\documentstyle[amsfonts,epsfig,cite]{JHEP}

\newcommand{\vb}[1]{{\mathbf{#1}}}
\newcommand{\r}[1]{\ref{#1}}
\newcommand{\lb}[1]{\label{#1}}
\newcommand{\lra}{\leftrightarrow}
\newcommand{\bc}{\begin{center}}
\newcommand{\ec}{\end{center}}
\newcommand{\be}{\begin{equation}}
\newcommand{\ee}{\end{equation}}
\newcommand{\bea}{\begin{eqnarray}}
\newcommand{\eea}{\end{eqnarray}}
\newcommand{\ba}[1]{\begin{array}{#1}}
\newcommand{\ea}{\end{array}}
\newcommand{\bz}{{\bar{z}}}

\newcommand{\bt}[1]{\begin{table}[h]\centering\begin{tabular}{#1}}
\newcommand{\et}[1]{\end{tabular}\caption{\small#1}\end{table}}
\newcommand{\fig}[3]{\EPSFIGURE[ht]{#1,width=163mm,origin=-100mm}{#2\label{#3}}}
\newcommand{\figh}[3]{\EPSFIGURE[h]{#1,width=163mm,origin=-100mm}{#2\label{#3}}}

\newcommand{\Tr}{{\mathrm{Tr}}\,}

\author{P. Castelo Ferreira\footnote{e-mail:
pcastelo@thphys.ox.ac.uk}\\Department of Physics -- Theoretical Physics, University of Oxford\\ 1
Keble Road, Oxford OX1 3NP, U.K.}
\author{Ian I.\ Kogan\footnote{e-mail: kogan@thphys.ox.ac.uk}\\
Department of Physics -- Theoretical Physics, University of Oxford\\ 1
Keble Road, Oxford OX1 3NP, U.K.}


\preprint{OUTP-00-31P\\
hep-th/0012188}

\title{Open and Unoriented Strings
from Topological Membrane\\
I. Prolegomena}

\abstract{We study open and unoriented strings in a Topological Membrane (TM)
theory through orbifolds of the bulk $3D$ space. This is achieved
by gauging discrete symmetries of the theory. Open and unoriented
strings can be obtained from all possible realizations of $C$, $P$ and 
$T$ symmetries. The important role of $C$ symmetry to distinguish between
Dirichlet and Neumman boundary conditions is discussed in detail.}

\begin{document}

\tableofcontents

\section{Introduction}
Although originally (and historically) open string theories were
considered as theories by themselves, it soon become evident that,
whenever they are present, they come along with closed (non-chiral)
strings. Moreover open string theories are obtained from closed
string theories by gauging certain symmetries of the closed theory
(see~\cite{SBH_00} and references therein for a discussion of
this topic). The way to get open strings
from closed strings is by gauging the world-sheet
parity~\cite{SBH_00,SBH_01,SBH_02}, $\Omega:z\rightarrow-\bz$. That is
we impose the identification $\sigma_2\cong-\sigma_2$, where
$z=\sigma_1+i\sigma_2$ and $\bz=\sigma_1-i\sigma_2)$
is the complex structure of the world-sheet manifold.
The spaces obtained in this way can be of two types: closed unoriented
and open oriented (and unoriented as well). These last ones are generally
called orbifolds and the singular points of the construction become boundaries.
The states (operators and fields of the theory in general) of the
open/unoriented theory are obtained from the closed oriented theory by
projecting out the ones which have negative eigenvalues of the parity
operator. This is obtained by building a suitable projection
operator $(1+\Omega)/2$ such that only the states of positive eigenvalues are
kept in the theory. Namely the identification 
$X^{I}(z,\bz)\cong X^{I}(\bz,z)$ or $X^{I}_{L}(z)\cong X^{I}_{R}(\bz)$
(in terms of the holomorphic and antiholomorphic parts of $X=X_L+X_R$)
holds.

Another construction in string theory is orbifolding the target space
of the theory under an involution of some symmetry of that space.
In this work we are going to consider only a $Z_2$ involution,
imposing the identification $X^{I}\cong-X^{I}$, where $X^{I}$
are the target space coordinates.
When combining both constructions, world-sheet and target space
orbifolding, we obtain open/unoriented theories in
orbifolds~\cite{SBH_03,SBH_04,SBH_05,SBH_05a}
or orientifolds ($X^{I}(z,\bz)=-X^{I}(\bz,z)$),
implying the existence of twisted sectors in the open/unoriented
theories.

Further to the previous discussion
both sectors (twisted and untwisted) need to be
present for each surface in order to ensure modular invariance of the
full partition function~\cite{SBH_00,SBH_06,SBH_07}.
One point we want to stress is that twisting in open strings can, for
the case of a $Z_2$ target space orbifold, be simply interpreted as
the choice of boundary conditions: Neumman or Dirichlet.

Toroidal compactification is an important construction in string
theories and in the web of target space dualities.
Early works considered also open string constructions in these
toroidal backgrounds~\cite{SBH_07,SBH_08}.  
In these cases we have
some compactified target space coordinates, say
$X^{J}(z+2\pi i,\bz-2\pi i)\cong X^{J}(z,\bz)+2\pi R$
($R$ is the radius of compactification of $X^{J}$),
the twisted states in the theory are the ones
corresponding to the points identified under
$X^{J}(z+2\pi i,\bz-2\pi i)\cong-X^{J}(z,\bz)+2\pi R$ or in
terms of the holomorphic and antiholomorphic parts of $X$
this simply reads $X_L^{I}(z)\cong-X_R^{I}(\bz)$.

An important result coming from these constructions is that the gauge
group of the open theory, the Chan-Paton degrees of freedom carried by 
the target space photon Wilson lines (only present in open theories) are
constrained, both due to dualities of open string theory~\cite{SBH_07}
and to modular invariance of open and unoriented
theories~\cite{SBH_07,SBH_08,SBH_09,SBH_10}. This will result in the
choice of the correct gauge group that cancels the anomalies in the theory.

One fundamental ingredient of string theory is modular invariance.
Although for bosonic string theory the constraints coming from genus 1
amplitudes are enough to ensure modular invariance at generic genus
$g$, it becomes clear that once the fermionic sector of superstring
theory is considered it is necessary to consider genus $2$ amplitude
constraints.  For closed strings (types II and 0) the modular group at
genus $g$ is $SPL(2g,{\mathbb{Z}})$ and the constraints imposed by
modular invariance at $g=2$ induce several possible projections in the
state space of the theory~\cite{MI_01,MI_02,MI_03,MI_04,MI_05} such
that the resulting string theories are consistent. Among them are the
well known GSO projections~\cite{GSO} that insure the correct
spin-statistics connection, project out the tachyon and ensure a
supersymmetric effective theory in the $10D$ target space.

Once we consider an open superstring theory (type I)
\textit{created} by orbifolding the world-sheet parities,
for each open (and/or unoriented) surface
a Relative Modular Group still survives the orbifold
at each genus $g$~\cite{SBH_06}.
Again in a similarly way to the closed theory the modular invariance under
these groups will result in generalized
GSO projections~\cite{SBH_06,SBH_06a,SBH_06b,SBH_06c}.

For a more recent overview of the previous topics
see~\cite{SBH_11,SBH_12} (see also~\cite{P_1} for an extensive
explanation of them).

The purpose of this work is to build open, open unoriented, and closed
unoriented string theories (with and without orbifolding of the target
space) from the Topological Membrane
(TM)~\cite{TM_00,TM_01,TM_02,TM_03,TM_04,TM_05,TM_06,TM_07,TM_08,TM_09,TM_10,TM_11,TM_12,TM_13,TM_14,TM_last}.
This approach consists of a Topological Massive Gauge Theory
(TMGT)~\cite{TMGT_00,TMGT_01,TMGT_02} living on a $3D$ membrane,
i.e. a Maxwell term and a gauge
Chern-Simons term, together with Topological Massive Gravity (TMG),
i.e. Einstein and a gravitational Chern-Simons term.  The membrane is
a $3D$ manifold $M=\Sigma\times[0,1]$ which has two boundaries
$\partial M=\Sigma_0+\Sigma_1$.  Gauge transformations induce chiral
Conformal Field Theories on the boundaries. The first works were
concerning only $3D$ pure Chern-Simons theories in the
bulk~\cite{W_1,MS_1,MS_2}.

Closed string theories are obtained as the effective boundary
theory, their world-sheet is the closed boundary $\partial M$.
Obtaining open string theory raises a problem, we need a open
world-sheet to define them. But the boundary of a
boundary is zero, $\partial \partial M=0$. So naively it seems that TM
cannot describe open strings since world-sheets are already a
boundary of a $3D$ manifold. The way out is to consider
orbifolding of the bulk theory. In this way the fixed points of the
orbifold play the role of the boundary of the $2D$ boundary of the $3D$
membrane. This proposal was first introduced by Horava~\cite{H_1} in
the context of pure Chern-Simons theories. We are going to extend his 
results to TMGT and reinterpret the orbifolded group as symmetries of
the full gauge theory.

Other works have developed Horava's idea. For a recent study on
WZNW orbifold constructions see~\cite{FS_1} (and references therein)
For an extensive study, although in a more formal way than our
work, of generic Rational Conformal Field Theories (RCFT)
with boundaries from pure $3D$ Chern-Simons theory
see~\cite{FS_2} (and references therein).
Nevertheless previously the monopole
processes were not studied. These are crucial for describing the
winding modes and T-duality in compact RCFT from the TM point of view
and, therefore, in compactified string theories.

We consider an orbifold of TM(GT) such that one new boundary is
created at the orbifold fixed point. To do this we gauge the discrete
symmetries of the $3D$ theory, namely $PT$ and $PCT$. Several $P$'s
are going to be defined as generalized parity operations. $C$ and $T$
are the usual $3D$ QFT charge conjugation and time inversion
operations (see~\cite{GD_1} for a review). The orbifolding of the
string target space corresponds in pure Chern-Simons membrane theory
to the quotient of the gauge group by a $Z_2$ symmetry~\cite{MS_1}.
As will be shown, in the full TM(GT), the discrete symmetry which
will be crucial in this construction is charge conjugation
$C$. Besides selecting between twisted and untwisted sectors in closed
unoriented string theory it will also be responsible for setting
Neumann and Dirichlet boundary conditions in open string theory. In
this work we are not going to consider more generic orbifold groups.

There are two main new ideas introduced in this work. Firstly the use
of all possible realizations of $P$, $C$ and $T$ combinations, which
constitute discrete symmetries of the theory, as the orbifold
group. Although the mechanism is similar to the one previously studied
by Horava for pure Chern-Simons theory, the presence of the Maxwell
term constrains the possible symmetries to $PT$ and $PCT$ type
only. Also the interpretation of the orbifold group as the discrete
symmetries in the quantum theory is new, as is the interpretation of
charge conjugation $C$ which selects between Neumman and Dirichlet
boundary conditions. This symmetry explains the T-duality of open
strings in the TM framework. It is a symmetry of the $3D$ bulk which
exchanges trivial topological configurations (without monopoles) with
non-trivial topological configurations (with monopoles). In terms of
the effective boundary CFT (string theory) this means exchanging
Kaluza-Klein modes (no monopole effects in the bulk) with winding
number (monopole effects in the bulk).

In section~\r{sec:riemman} we start by introducing genus 0 (the
sphere), and genus 1 (the torus), Riemann surfaces and their possible
orbifolds under discrete symmetries which we identify with generalized
parities $P$.

Section~\r{sec:cft.bc} gives an account of Neumann and Dirichlet
boundary conditions in usual CFT using the Cardy method~\cite{CFT_3}
of relating $n$ point full correlation functions in boundary
Conformal Field Theory with $2n$ chiral correlation functions in the
theory without boundaries.

Then, in section~\r{sec:tmgt} we give a brief overview of the discrete
symmetries of $3D$ QFT and use it to orbifold TM(GT). We enumerate the
$3D$ configurations compatible with the several orbifolds,
both at the level of the field configurations and of the particular
charge spectrums corresponding to the resulting theories.
It naturally emerges from the $3D$ membrane that the configurations
compatible with $PCT$ correspond to Neumann boundary conditions (for
open strings) and to untwisted sectors (for closed unoriented).
The configurations compatible with $PT$ correspond to
Dirichlet boundary conditions (for open strings) and twisted sectors
(for closed unoriented). The genus 2 constraints are discussed here
although a more detailed treatment is postponed for future work.
Further it is shown
that Neumann (untwisted) corresponds to the absence of monopole
induced processes while for Dirichlet (twisted) these processes play
a fundamental role. A short discussion on
T-duality show that it has the same bulk meaning as modular
invariance, they both exchange $PT\lra PCT$.

\section{\lb{sec:riemman}Riemann Surfaces:\\ from Closed Oriented to
Open and Unoriented}
Any open or unoriented manifold $\Sigma_u$ can, in general, be obtained
from some closed orientable manifolds  $\Sigma$ under identification
of a $Z_2$ (or at most two $Z_2$) involution
\be
\ba{cccl}
\pi:&\Sigma&\rightarrow&\Sigma_u=\Sigma/Z_2\\
    &(x,-x)&\rightarrow&x
\ea
\lb{proj}
\ee
such that each point in $\Sigma_u$ has exactly two corresponding
points in $\Sigma$ conjugate in relation to the $Z_2$ involution(s).
The pair $(x,-x)$ in the last equation is symbolic, the second
element stands for the action of the group $Z_2$, $z_2(x)=-x$,
in the manifold. Usually this operation is closely related with
parity as will be explained bellow.
Although in this work our perspective is that we start from a full
closed oriented theory and orbifold it, there is the 
reverse way of explaining things. This means that any theory
defined in an open/unoriented manifold is equivalently defined
in the closed/oriented manifold which \textit{doubles} (consisting of
two copies of) the original open/unoriented.

Let us summarize how to obtain the disk $D_2$ (open orientable) and
projective plane $RP_2$ (closed unorientable) out of the sphere $S^2$ and
the annulus $C_2$ (open orientable), the M\"{o}bius Strip (open
unorientable) and Klein bottle $K_2$ (closed unorientable)
out of the torus $T^2$.

\subsection{The Projective Plane and the Disk obtained from the Sphere}

For simplicity we choose to work in complex stereographic coordinates
\mbox{$(z=x_1+ix_2$}, $\bz=x_1-ix_2)$ such that the sphere is identified with
the full complex plane.
The sphere has no moduli and the Conformal Killing Group (CKG) is $PSL(2,\mathbb{C})$.
A generic element of this group is $(a,b,c,d)$ with the restriction 
$ad-bc=1$. It acts in a point as
\be
z'=\frac{a z+b}{c z+d}
\lb{PSLC}
\ee
It has then six real parameters, that is, six generators. That is to say that
the sphere has six Conformal Killing Vectors (CKV's).
It is necessary to use two coordinate charts to cover the full sphere, 
one including the north pole and the other one including the south
pole. Usually it is enough to analyze the theory defined on the sphere
only for one of the patches but it is necessary to check that the
transformation between the two charts is well defined. In
stereographic complex coordinates the map between the two charts
(with coordinates $z,\bz$ and $u,\bar{u}$) is given by $z\to1/u$ and
$\bz\to 1/\bar{u}$.

The {\bf disk $D_2$} can be obtained from the sphere under the identification
\be
z\cong\bz
\lb{D2_id}
\ee
This result is graphically pictured in figure~\r{fig_D2} and
consists in the involution  of the manifold $S^2$ by the group $Z^{P_1}$, $D_2=S^2/Z^{P_1}$.
There are one boundary corresponding to the real line in the complex
plane and the disk is identified with the upper half complex plane.
\fig{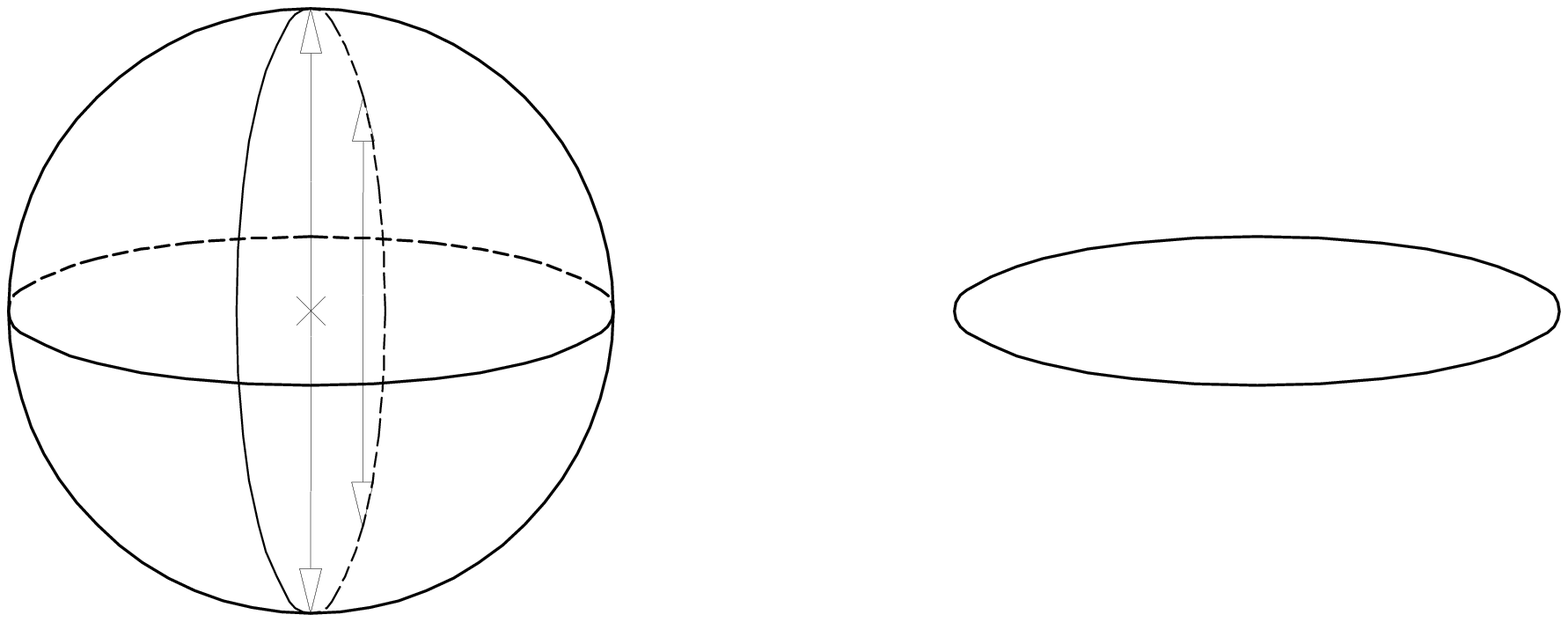}{The disk $D_2=S^2/P_1$ obtained from the sphere under
the involution given by the parity operation $P_1$.}{fig_D2}

It is straightforward to see that the non trivial element of
$Z^{P_1}$ is nothing else than the usual $2D$ parity transformation
\be
\ba{cccc}
P_1:&z      &\rightarrow&\bz\vspace{.1 cm}\\
       &\bz    &\rightarrow&z
\ea
\lb{P}
\ee
The CKG of the disk is the subgroup of $PSL(2,\mathbb{C})$ which maintains
constraint (\r{D2_id}), that is $PSL(2,\mathbb{R})$.

From the point of view of the fields defined in the sphere this corresponds
to the usual $2D$ parity transformation. In order that the theory be well
defined in the orbifolded sphere we have to demand the fields of the
theory to be compatible with the construction
\be
\ba{rcl}
f(z)&=&f(P_1(z))\vspace{.1 cm}\\
\phi_i[x_j]&=&P_1\phi_i[P_1(x_j)]
\ea
\lb{Of}
\ee
where the first equation applies to scalar fields and the second to
vectorial ones. For tensors of generic dimensions $d$ (e.g. the metric
or the antisymmetric tensor) the transformation is easily generalized
to be $T(x)=P_1^dT(P_1(x))$.

In order to orbifold the theory defined on the sphere we can
introduce the projection operator
\be
P_{1,\mathrm{proj}}=\frac{1}{2}(1+P_1)
\ee
which projects out every operator with odd parity eigenvalue and
keeps in the theory only field configurations compatible with the
$Z_2$ involution.

To obtain the {\bf projective plane $RP_2$} we need to make the identification
\be
z\cong -\frac{1}{\bz}
\lb{sphere_id}
\ee
This result is graphically pictured in figure~\r{fig_RP2} and again is
an involution of the sphere
$RP_2=S^2/Z_2^{P_2}$.
The resulting space has no boundary
and no singular points. But it is now an unoriented manifold.
\fig{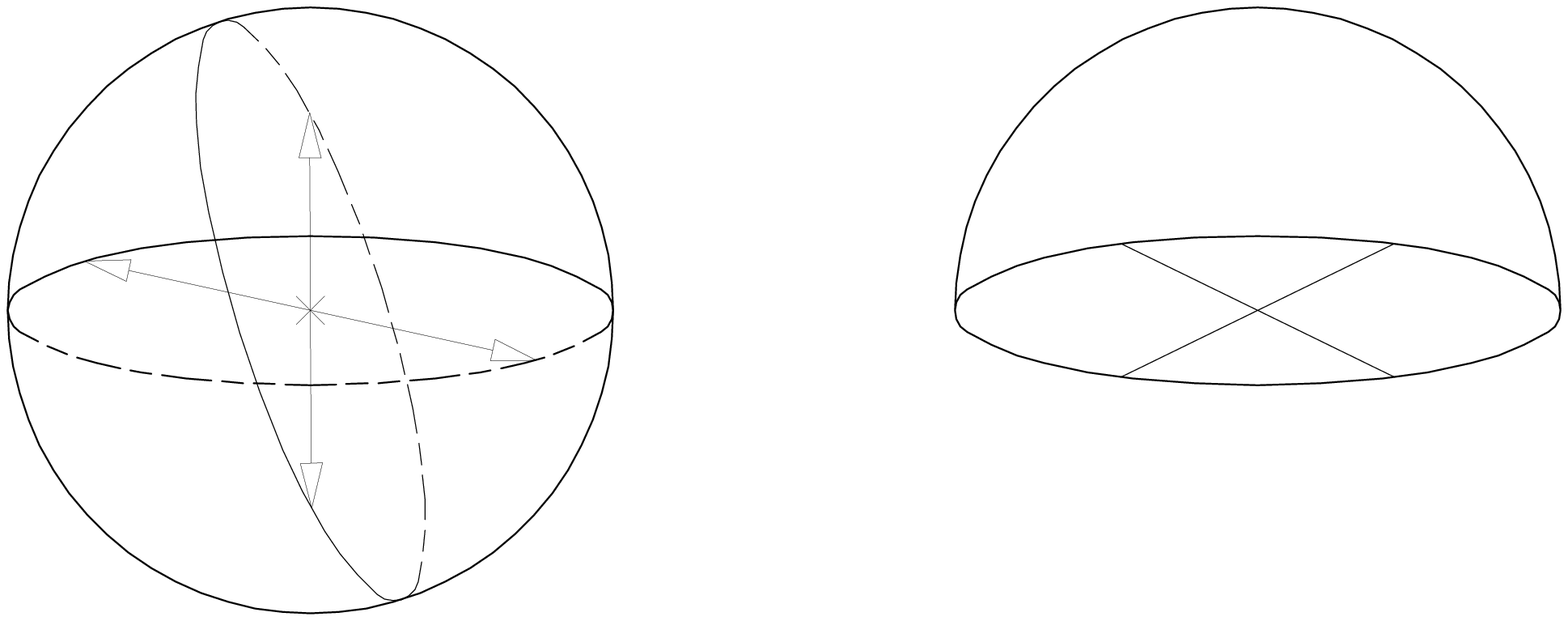}{The projective plane $RP_2=S^2/P_2$ obtained from the sphere
under the involution given by the parity operation $P_2$.}{fig_RP2}

This identification can be thought of as two operations. The action of the element
$\alpha=(0,-1,1,0)\in Z_2^\alpha\subset SL(2,{\mathbb{C}})$ followed
by the operation of parity as given by~(\r{P}). Note that $\alpha(z)=-1/z$ but
$P_1\alpha(z)=-1/\bz$ as desired.
In this case we can define a new parity operation
$P_2\in Z_2^{P_2}=Z_2^{P_1}\times Z_2^\alpha$ as
\be
\ba{rrcl}
P_2:&z  &\rightarrow&\displaystyle-\frac{1}{\bz}\vspace{.1 cm}\\
  &\bz&\rightarrow&\displaystyle-\frac{1}{z}
\ea
\lb{PP}
\ee
From the point of view of the fields defined in the sphere we could
use the usual parity transformation since any theory defined in the
sphere should be already invariant under transformation (\r{PSLC})
such that $PSL(2,\mathbb{C})$ is a symmetry of the theory.
But in order to have a more transparent picture
we use the definition (\r{PP}) of $P_2$ and demand that
\be
\ba{rcl}
f(z)&=&f(P_2(z))\\
\phi_i[x_j]&=&P_2\left(\,\phi_i[P_2(x_j)]\,\right)
\ea
\lb{Pf}
\ee
where the first equation concerns to scalar fields and the second to
vectorial ones. For tensors of generic dimensions $d$ (as the metric
or the antisymmetric tensor) the transformation is again easily
generalized to be $T(x)=P_2^dT(P_2(x))$.

The CKG is now $SO(3)$, the usual rotation group. It is the subgroup of
$PSL(2,{\mathbb{C}})/Z_2^\alpha$ that maintains constraint~(\r{D2_id})

\subsection{\lb{sec:torus_inv}The annulus, M\"{o}bius strip and Klein bottle from the Torus}

Let us proceed to genus one closed orientable manifold, the torus.
It is obtained from the complex plane under the identifications
\be
z\cong z+2\pi\cong z+2\pi (\tau_1+i\tau_2)
\ee
There are two modular parameters $\tau=\tau_1+i\tau_2$ and two CKV's.
The action of the CKG, the translation group in the
complex plane, is
\be
z'=z+a+ib
\ee
with $a$ and $b$ real.
The metric is simply $\left|dx^1+\tau dx^2\right|$ and the
identifications on the complex plane are invariant under the two
operations
\be
{\mathcal{T}}:\tau'=\tau+1\ \ \ \ \ \ \ \ \ \ \ {\mathcal{S}}:\tau'=-\frac{1}{\tau}
\ee
These operations constitute the \textbf{modular group}
$PSL(2,{\mathbb{Z}})$. That is
\be
\tau'=\frac{a\tau+b}{c\tau+d}
\ee
with $a,b,c,d\in \mathbb{Z}$ and $ad-bc=1$.

The {\bf annulus $C_2$} (or topologically equivalent, the cylinder) is
obtained from the torus
with $\tau=i\tau_2$ under the identification
\be
z\cong-\bz
\lb{C2_id}
\ee
This result is symbolically picture in figure~\r{fig_C2}.
\fig{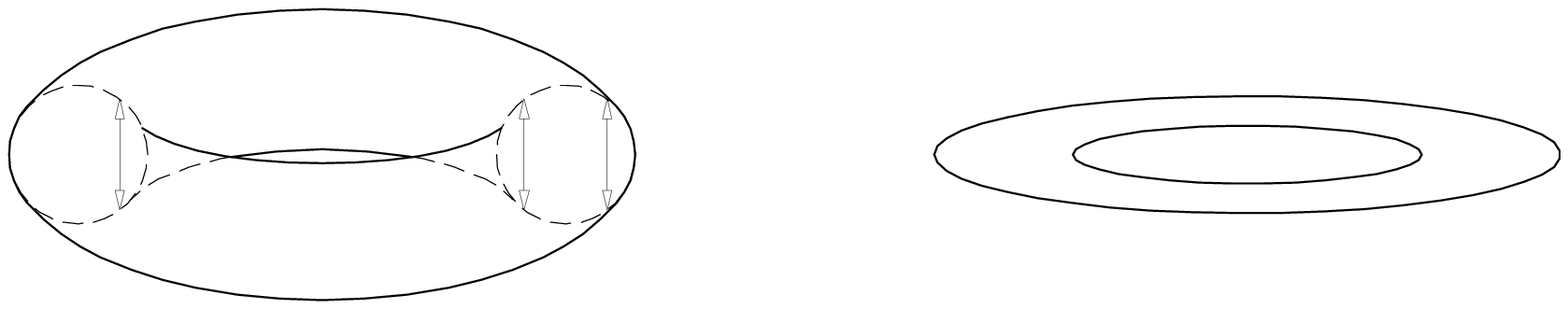}{The annulus (or cylinder) $C_2=T^2/\Omega$ obtained
from the torus under the involution given by the parity
operation $\Omega$.}{fig_C2}

There is now one modular parameter $\tau_2$ and no modular
group. There is only one CKV being the CKG action given by $z'=z+ib$,
translation in the imaginary direction.
In terms of the fields defined in the torus this correspond to the
projection under the parity operation
\be
\ba{rrcl}
\Omega:&z      &\rightarrow&-\bz\vspace{.1 cm}\\
       &\bz    &\rightarrow&-z
\ea
\lb{omega}
\ee

The {\bf M\"{o}bius strip $M_2$} can be obtained from the annulus
(obtained from the torus with $\tau=2i\tau_2$) by the
identification under the element $\tilde{a}$~\cite{P_1} of
the translation group
\be
\tilde{a}:z\to z+2\pi\left(\frac{1}{2}+i \tau_2\right)
\lb{M2_id}
\ee
Note that $\tilde{a}$ belongs to the translation group of the torus,
not of the disk, and that $\tilde{a}^2=1$. This construction
corresponds to two involutions, so the orbifolding group
is constituted by two $Z_2$'s,
$M_2=T^2/(Z_2^{\Omega}\subset\hspace{-12.9pt}{\times}
Z_{2}^{\tilde{a}})$, where $\subset\hspace{-12.9pt}{\times}$ stands
for the semidirect product of groups. Thus the ratio of areas between the
M\"{o}bius strip and the original torus is $1/4$ contrary to the $1/2$
of the remaining open/unoriented surfaces obtained from the torus, due
to the extra projection operator $(1+\tilde{a})/2$ taking from the
annulus to the strip.

In terms of the fields living on the torus we can think of this
identification as the projection under a new discrete symmetry, which
we also call parity
\be
\tilde{\Omega}\equiv\tilde{a}\circ\Omega
\lb{omegat}
\ee
Although this operation does not seem to be a conventional parity
operation note that, applying it twice to some point, we retrieve
the same point, $\tilde{\Omega}^2=1$. It is in this sense a generalized
parity operation.

\figh{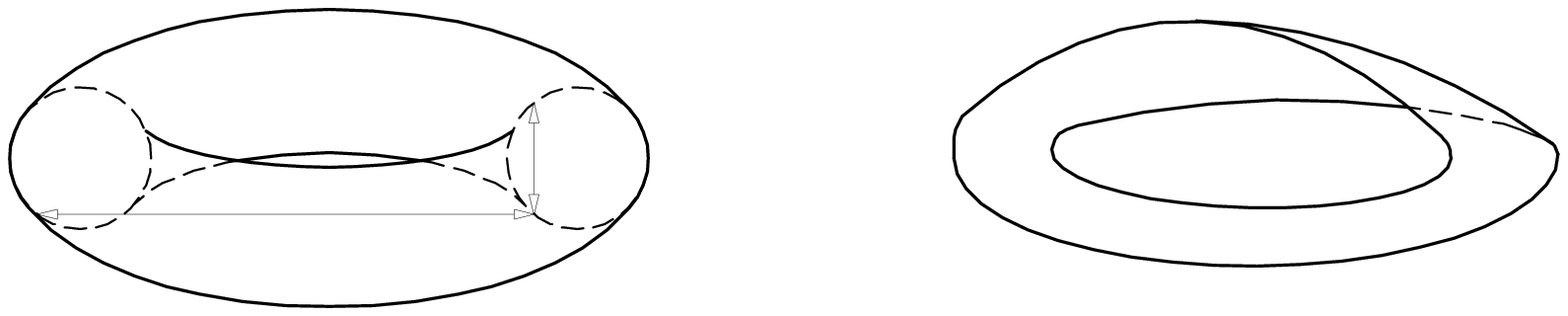}{The M\"{o}bius strip $M_2=T^2/\tilde{\Omega}$ obtained
from the torus under the involution given by the parity
operation $\tilde{\Omega}$.}{fig.M2}

The previous construction is presented, for example, in Polchinski's
book~\cite{P_1}. Let us note however that
one can build the M\"{o}bius strip directly from a torus~\cite{SBH_00}
with moduli $\tau=1/2+i\tau_2$ under the involution by $\Omega$ as given
in~(\r{omega})~\footnote{The authors thank 
the referee for this useful remark!}.
In this case the ratio of areas between the original torus and the
involuted surface is $1/2$ as the other
involutions studied in this section. As we will show later
both constructions correspond to the same region on the
complex plane. The first one results from two involutions
of a torus ($\tau=2i\tau$) with double the area
of the second construction ($\tau=i\tau$). In this
sense both constructions are equivalent.
The M\"{o}bius strip orbifolding is pictured in figure~\r{fig.M2}.

Again there is one modular parameter $\tau_2$ and no modular
group. The only CKV is again the translation in the imaginary direction.

The {\bf Klein bottle $K_2$} is obtained from the torus with $\tau=2i \tau_2$ under the identification
\be
z\cong-\bz+2\pi i \tau_2
\lb{torusid}
\ee
This result is pictured in figure~\r{fig.K2}.
\figh{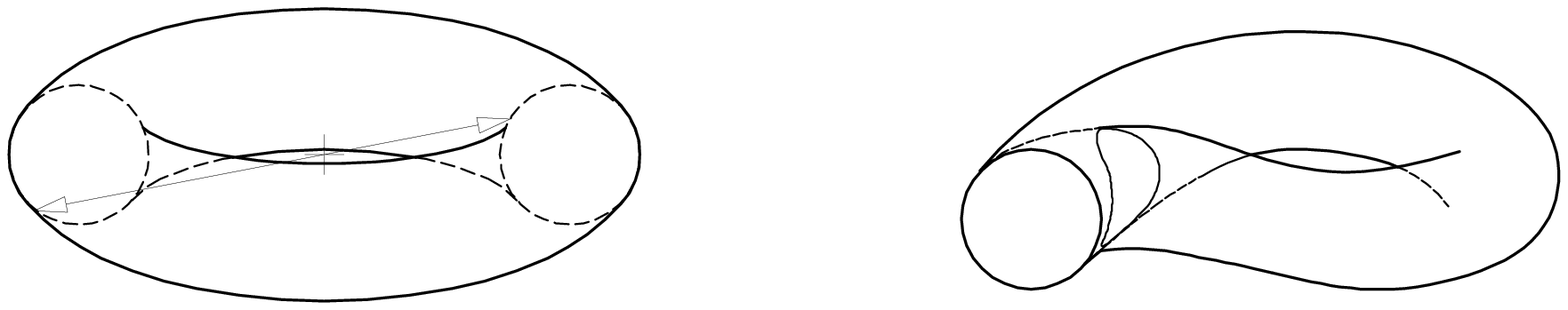}{The Klein bottle $K_2=T^2/\Omega'$ obtained from the
torus under the involution given by the parity operation $\Omega'$.}{fig.K2}

\bt{|cccc|cccc|}
\hline$\vb{S_2}$&$P_1$  &$P_2$          &\ \ \ \ &$\vb{T_2}$ &$\Omega$  &$\tilde{\Omega}$                       &$\Omega'$\\\hline
&&&&&&&\\
        &$z\lra\bz$&$z\lra-1/\bz$&        &        &$z\lra-\bz$&$\tilde{a}\circ\Omega$&$z\to-\bz+2\pi i\tau_2$\\
        &          &             &        &        &           & &$\bz\to-z+2\pi i\tau_2$\\
&&&&&&&\\
$S^2/P=$&$D_2$     &$RP_2$       &        &$T^2/P=$&$C_2$      &$M_2$                      &$K_2$\\
C/O&O/O     &C/U          &        &C/O     &O/O        &O/U&C/U\\
&&&&&&&\\\hline
\et{\lb{tabpar}Parity operations for the topology $T^2\times[0,1]$.
The torus geometry considered is $\tau=i\tau_2$ for $\Omega$ and
$\tau=2 i\tau_2$ for $\tilde{\Omega}$ and $\Omega'$. Note that $M_2$
can also be obtained from the torus with $\tau=1/2+i\tau_2$
considering the parity $\Omega$.
In the labels of the last line the first letter stands for {\bf O}pen
or {\bf C}lose surface while the second letter stands for
\textbf{O}riented or \textbf{U}noriented.}

The bottle is the involution of the torus $K_2=T^2/Z^{\Omega'}_2$,
has one parameter CKG with one CKV, translations in the imaginary
direction. There is one modulus $\tau_2$ and no modular group.
The resulting manifold has no boundary and no singular points
but is unoriented.

Again we can define a new parity transformation $\Omega'$
\be
\ba{cccc}
\Omega':&z  &\rightarrow&\displaystyle-\bz+2\pi i\tau_2\vspace{.1 cm}\\
          &\bz&\rightarrow&\displaystyle -z+2\pi i\tau_2
\ea
\lb{omega'}
\ee

We summarize in table~\r{tabpar} all the parity operations we have just
studied together with the resulting involutions (or orbifolds).

\section{\lb{sec:cft.bc}Conformal Field Theory -\\ Correlation Functions and Boundary Conditions}

To study string theory we need to know the world-sheet CFT. In a
closed string theory they are given by CFT on a closed Riemann
surface, the simplest of them is the sphere, or equivalently the
complex plane. To study open strings we need to study CFT on open
surfaces. As was shown by Cardy~\cite{CFT_3} n-point correlation
functions on a surface with a boundary are in one-to-one
correspondence with chiral $2n$ point correlation functions on the
double surface\footnote{One of the constructions presented to obtain
the M\"{o}bius strip uses the double involution under $\tilde{\Omega}$.
In that case $n$ insertions on it correspond to $4n$ in the original
torus.} (for more details and references see~\cite{FMS}).

We will study the disk and the annulus, so we double the number
of charges (vertex operators) by inserting charges $\pm q$ (vertex operators
with $\Delta=2q^2/k$) in the Parity conjugate points. Note that the
sign of the charges inserted depends on the type of boundary conditions
that we want to impose but the conformal dimension of the
corresponding vertex operator is the same.

We summarize the 2, 3 and 4-point holomorphic correlation functions of
vertex operators for the free boson
\be
\ba{rcl}
<\phi(z_1)\phi(z_2)>&=&z_{12}^{-2\Delta}\vspace{.1 cm}\vspace{.1 cm}\\
<\phi(z1)\phi(z2)\phi(z3)>&=&z_{12}^{-\Delta_1-\Delta_2+\Delta_3}z_{13}^{-\Delta_1+\Delta_2-\Delta_3}z_{23}^{\Delta_1-\Delta_2-\Delta_3}\vspace{.1 cm}\vspace{.1 cm}\\
<\phi(z_1)\phi(z_2)\phi(z_3)\phi(z_4)>&=&\prod_{i<j}z_{ij}^{2q_iq_j/k}
\ea
\lb{correl}
\ee
where in all the cases $\sum q_i=0$, otherwise they vanish.

\subsection{Disk}

We will take the disk as the upper half complex plane. As explained before
it is obtained from the sphere (the full complex plane) by identifying
each point in the lower half complex plane with it's conjugate in the
upper half complex plane. In terms of correlation functions
\be
\left<\phi_q(x,y)\right>_{D_2}=\left<\phi_q(z)\phi_{-q}(\bz)\right>_{S^2}
\ee
where we replaced $z=x+iy$ in the the first equation of~(\r{correl}),
$y$ is the distance to the real axis while $x$ is taken to be the
horizontal distance (parallel to the real axis) between vertex
insertions.

\subsubsection{Dirichlet Boundary Conditions}

As it is going to be shown, when the mirror charge have opposite
sign the boundary conditions are Dirichlet.

The 2-point correlation
function restricted to the upper half plane is simply the expectation value
\be
\left<\phi_q(x,y)\right>=\frac{1}{(2y)^{2\Delta}}
\lb{1pD}
\ee
Insertion of vertex operators
(from the unity) in the boundary is not compatible with the boundary
conditions since the only charge that can exist there is $q=0$ (since
$q=-q=0$ in the boundary).
Taking the limit $y\to 0$ the expectation value (\r{1pD}) blows up but
this should not worry us, near the boundary the two charges \textit{annihilate}
each other. This phenomena is nothing else
than the physical counterpart of the operator fusion rules
$\phi_q(y)\phi_{-q}(-y)\to(2y)^{-2\Delta}\phi_0(y)$. That is
$\left<\phi_0\right>_{\partial D_2}=\left<1\right>_{\partial D_2}$
in the boundary of the disk. 

3-point correlation functions cannot be used for the same reason, one
of the insertions would need to lie in the boundary but that would
mean $q_3=0$, the other two charges had to be inserted symmetrically
in relation to the real axis and would imply $q_1=-q_2$.
This reduces the 3-point correlator to a 2-point one in the full plane.

For 4-point vertex insertions consider $q_1$ and $q_3$ in the upper
half plane, $q_2$ (inserted symmetrically to $q_1$) and $q_4$ (inserted
symmetrically to $q_3$) in the lower half plane. As pictured
in figure~\r{fig.confd2} the most generic  configurations is
$q_1=-q_2=q$ and $q_3=-q_4=q'$. Making $z_2=\bz_1=-iy$
and $z_4=\bz_3=x-iy'$ we obtain the corresponding 2-point correlators
in the upper half plane
\be
\left<\phi_q(0,y)\phi_{q'}(x',y')\right>=\frac{1}{(2y)^{2\Delta}(2y')^{2\Delta'}}
\left(1-\frac{4yy'}{x^2+(y+y')^2}\right)^{\frac{2qq'}{k}}
\ee

Again note that we cannot insert boundary operators without
changing the boundary conditions.
In the limit $x\to\infty$ both correlators behave like
\be
\lim_{x\to\infty}\left<\phi(y_1)\phi(y_2)\right>= \frac{1}{(4y_1y_2)^{2\Delta}}
\ee

When we approach the boundary the correlators go to infinite
independently of the value of $x$. This fact can be explained by the
kind of boundary conditions we are considering, they are such that
when the fields approach the boundary they become infinitely correlated
independently of how far they are from each other.
Therefore this must be Dirichlet boundary conditions,
the fields are fixed along the boundary, furthermore, as stated before
their expectation value is $\left<1\right>$. It doesn't mater
how much apart they are, they are always correlated on the boundary.
The tangential derivative to the boundary of the expectation value
$\partial_x\left.\left<\phi\right>\right|_{\partial D_2}=0$
also agrees with Dirichlet boundary conditions.

\subsubsection{Neumann Boundary Conditions}

For the case of the mirror charge having the same sign of the original
one the boundary conditions will be Neumann. The expectation value for
the fields in the bulk vanishes since the 2-point function
$\left<\phi_q(z_1)\phi_q(z_2)\right>=0$ in the full plane.
Nevertheless we can evaluate directly the non-zero 2-point correlation
function in the boundary
\be
\left<\phi_q(0)\phi_{-q}(x)\right>=\frac{1}{x^{2\Delta}}
\lb{2pN}
\ee
Note that contrary to the previous
discussion, concerning Dirichlet boundary conditions, in this case
$q\neq 0$ on the boundary since the mirror charges have the same sign
and the correlation function vanishes in the limit $x\to\infty$ indicating that
the boundary fields become uncorrelated.

The 3-point correlation function in the full plane must be considered
with one charge $-2q$ in the boundary and two other charges $q$
inserted symmetrically in relation to the real axis (see
figure~\r{fig.confd2}).
In the upper half plane this corresponds to one charge insertion in
the boundary and one in the bulk
\be
\left<\phi_{-2q}(0,0)\phi_q(x,y)\right>=\left(\frac{2y}{x^2+y^2}\right)^{2\Delta}
\lb{3pN}
\ee
Note that in the limit $y\to 0$ the fusion rules apply and we
obtain~(\r{2pN}) with $\Delta$ replaced by $4\Delta$.

For the 2-point function in the upper plane we have to consider the
4-point correlation function in the full plane with
$q_1=q_2=-q_3=-q_4=q$, where $q_2$ is inserted symmetrically to $q_1$
in relation to the real axis and $q_4$ to $q_3$. We obtain the bulk correlator
\be
\left<\phi_{q}(0,y)\phi_{-q}(x,y')\right>=\left(\frac{4yy'}{x^2(x^2+(y+y')^2)}\right)^{2\Delta}
\lb{4pN}
\ee
\fig{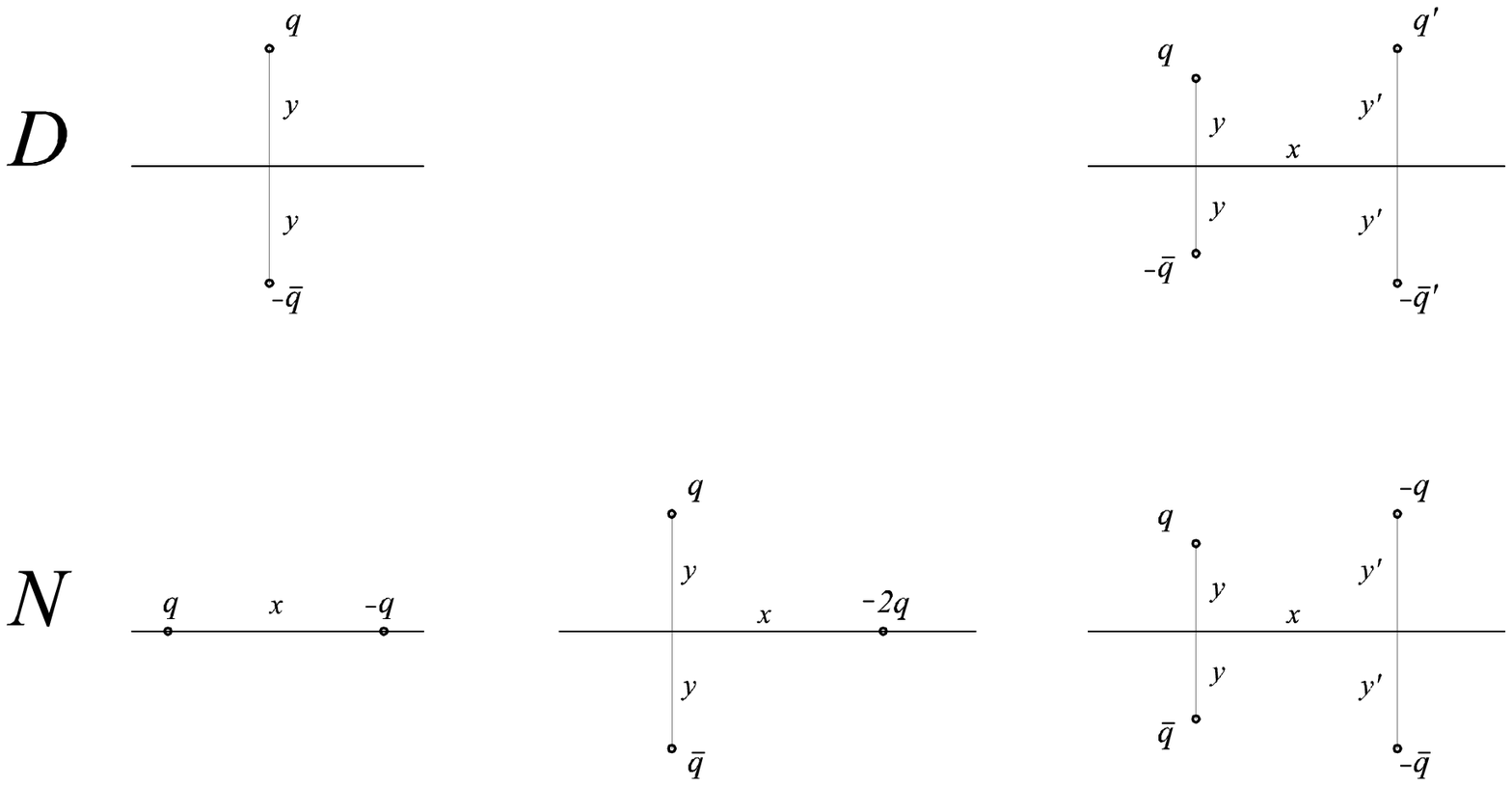}{The several possible analytical continuations of the disk (the
upper half plane) to the sphere (full plane) holding 2, 3 and 4-point
correlation functions for Neumann (N) and Dirichlet (D) boundary
conditions.}{fig.confd2}

Again in the limit $x\to\infty$ this correlator vanishes. This corresponds to
Neumann boundary conditions. The normal derivative to the boundary
of~(\r{4pN}) vanishes on the boundary
$\partial_y\left.\left<\phi(0)\phi(x)\right>\right|_{\partial D_2}=0$.

For the case of one compactified free boson the process follows in
quite a similar way. The main difference resides in the fact that the
right and left spectrum charges are different. Taking a charge
$q=m+kn/4$ its image charge is now $\pm\bar{q}$,
where $\bar{q}=m-kn/4$. In this way we have to truncate the
spectrum holding $q=-\bar{q}=kn/4$ for Dirichlet boundary conditions, 
and $q=\bar{q}=m$ for Neumann boundary conditions, in a pretty
similar way as it happens in the Topological Membrane. We summarize in
figure~\r{fig.confd2} the results derived here.

\subsection{Annulus}

We consider the annulus to be a \textit{half} torus.
For simplicity we take the torus to be the region of the complex plane
$[-\pi,\pi]\times[0,2\pi\tau]$ (and the annulus the region
$[0,\pi]\times[0,2\pi\tau]$).
We use $z=x+iy$ with $x\in [-\pi,\pi]$ and $y\in [0,2\pi\tau]$. Here
$y$ is the vertical distance (parallel to the imaginary axis) between
vertex insertions while $x$
is taken to be the distance to the imaginary axis.

\subsubsection{Dirichlet Boundary Conditions}

Considering mirror charges with opposite sign, 2-point correlations in 
the torus correspond to the bulk expectation value in the annulus
\be
\left<\phi(x,y)\right>=\frac{1}{(2x)^{2\Delta}}
\ee
As in the case of the disk, it blows up in the boundary. But in the
boundary this correlation function is not valid since the two charges
annihilate each other. Therefore the only possible charge insertions in 
the boundary are $q=0$, that is the identity operator.

Again 3-point correlation functions cannot be used in this case.

For 4-point vertex insertion consider $q_1$ and $q_3$ inserted to the
right of the imaginary axis and $q_2$ and $q_4$ their mirror
charges. The most generic configuration is $q_1=-q_2=q$ and
$q_3=-q_4=q'$ with $z_1=-\bz_2=x$ and $z_3=-\bz_3=x'+iy$.
We obtain the 2-point correlation function in the annulus
\be
\left<\phi_q(x,0)\phi_{q'}(x',y)\right>=\frac{1}{(2x)^{2\Delta}(2x')^{2\Delta'}}
\left(1-\frac{4xx'}{(x+x')^2+(y)^2}\right)^{2qq'/k}
\ee

Again the same arguments used for the disk apply. There cannot exist
boundary insertions other than the identity and the tangential
derivative to the boundary
$\partial_y\left.\left<\phi\right>\right|_{\partial C_2}=0$ vanish.

\subsubsection{Neumann Boundary Conditions}

Considering now the mirror charges having the same sign, again the
fields in the bulk have zero expectation value. But the 2-point
boundary correlation function is computed to be
\be
\left<\phi_q(0,0)\phi_{-q}(\pi,y)\right>=\frac{1}{\pi^{2\Delta}+y^{2\Delta}}
\ee
where we take one insertion in each boundary. In the case that the
insertions are in the same boundary the factor of $\pi^{2\Delta}$ is
absent.

The 3-point function in the torus corresponds either to 2-point
function in the annulus (taking only one insertion in the boundary) or 
to 3-point function (taking all the insertions in the boundaries).
Taking one insertion in the bulk $\phi_q(x,0)$ (with mirror image
$\phi_q(-x,0)$) and other in the boundary $\phi_{-2q}(\pi,y)$ we obtain
\be
\left<\phi_q(x,0)\phi_{-2q}(\pi,y)\right>=\left(\frac{x}{\pi^2+x^{2}+y^{2}}\right)^{2\Delta}
\ee
If the insertion is the boundary $x=0$ the factor of $\pi^2$ is
absent.

The 4-point function in the torus corresponds in the annulus either to 
a2-point
function (bulk insertions), 3-point function (two vertices in the
boundaries) or 4-point function (all vertices in the boundary).
Taking all vertex insertions in the bulk as pictured in
figure~\r{fig.confc2} we obtain
\be
\left<\phi_q(x,0)\phi_{-q}(x',y)\right>=\left(\frac{4xx'}{((x-x')^{2}+y^{2})((x+x')^{2}+y^{2})}\right)^{2\Delta}
\ee
\fig{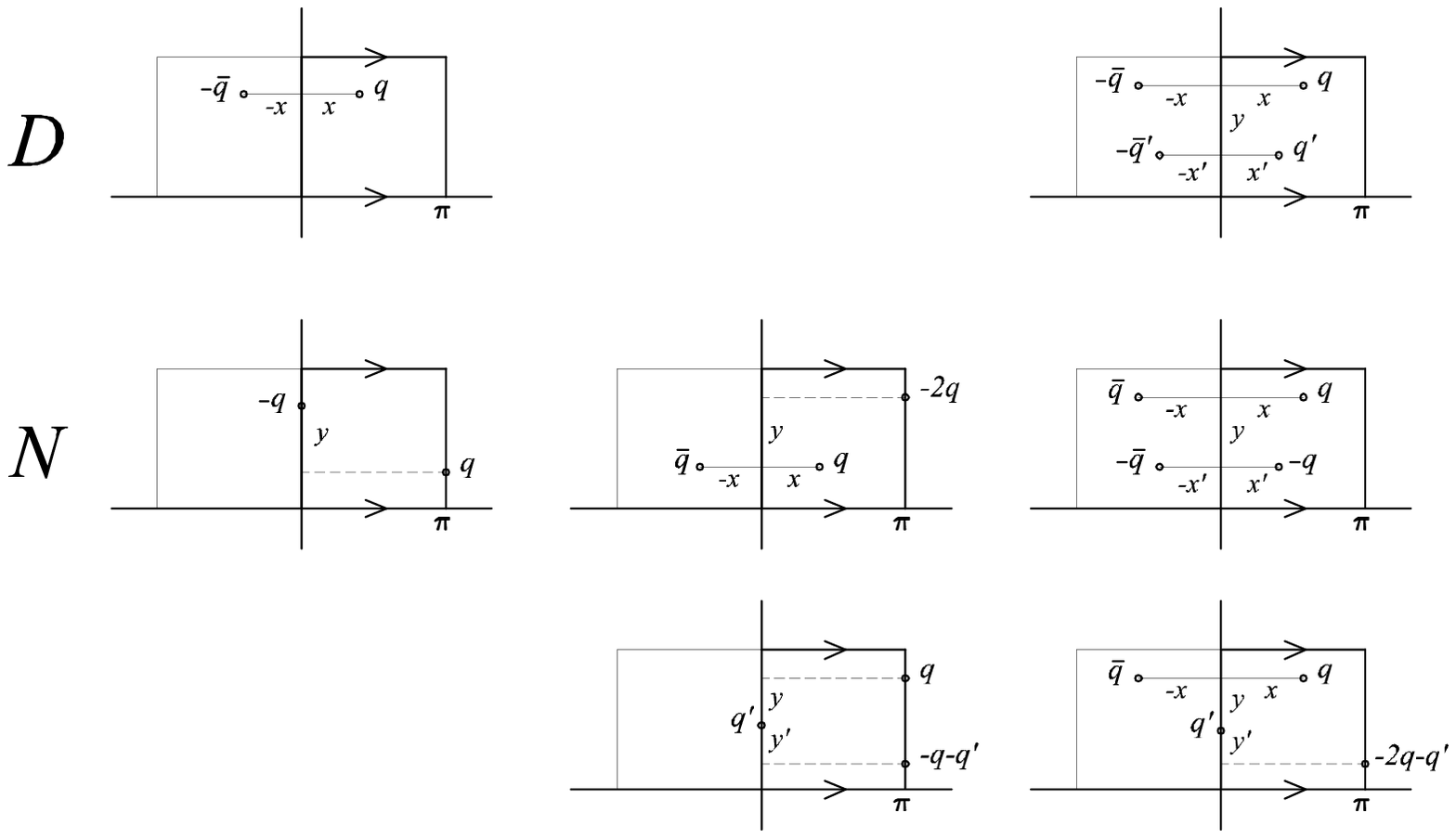}{The several possible analytical continuations of the annulus
to the torus holding 2, 3 and 4-point
correlation functions, for Neumann (N) and Dirichlet (D) boundary
conditions.}{fig.confc2}
As an example of two insertions in the boundaries take them to be both 
in the boundary $x=0$, we obtain
\be
\left<\phi_q(x,0)\phi_{q'}(0,y)\phi_{-q-q'}(0,y+y')\right>=\frac{1}{y^{2\Delta'}}\left(\frac{2x}{(x^2+y'^2)^2}\right)^{2\Delta}\left(\frac{x^2+(y+y')^2}{y^2(x^2+y'^2)}\right)^{2qq'/k}
\ee

We can stop here, for our purposes it is not necessary to exhaustively 
enumerate all the possible cases. As expected the normal derivative to 
the boundary of these correlation functions
($\partial_x\left<\ldots\right>$) vanishes at the boundary.
These results are summarized in figure~\r{fig.confc2}.

For the case of one compactified free boson the process follows as
explained before. The spectrum must be truncated
holding $q=-\bar{q}=kn/4$ for Dirichlet boundary conditions 
and $q=\bar{q}=m$ for Neumann boundary conditions.

\section{\lb{sec:tmgt}TM(GT)}
Is now time to turn to the $3D$ TM(GT). In this section we present
results derived directly from the bulk theory and its properties. The
derivations of the results presented here are in agreement with the CFT
arguments in the last section.

Take for the moment a single compact $U(1)$ TMGT corresponding
to $c=1$ CFT with action
\be
S=\int_{M}dtd^2z\left[-\frac{\sqrt{-g}}{\gamma}F_{\mu\nu}F^{\mu\nu}+\frac{k}{8\pi}\epsilon^{\mu\nu\lambda}A_\mu F_{\nu\lambda}\right]
\label{S}
\ee
where $M=\Sigma\times[0,1]$ has two boundaries $\Sigma_0$ and
$\Sigma_1$. $\Sigma$ is taken to be a compact manifold,
$t$ is in the interval $[0,1]$ and $(z,\bz)$ stand for complex
coordinates on $\Sigma$. From now on we will use them by default.

As widely known this theory induces new degrees of freedom in the
boundaries, which are fields belonging to $2D$ chiral CFT's theories
living on $\Sigma_0$ and $\Sigma_1$.

The electric and magnetic fields are defined as
\be
\ba{rcl}
E^i&=&\displaystyle\frac{1}{\gamma}F^{0i}\vspace{.1 cm}\\
B  &=&\partial_z A_\bz-\partial_\bz A_z
\ea
\ee
and the Gauss law is simply
\be
\partial_iE^i+\frac{k}{4\pi}B=\rho_0
\ee

Upon quantization the charge spectrum is
\be
Q=m+\frac{k}{4}n
\lb{charge}
\ee
for some integers $m$ and $n$.
Furthermore it has been proven in~\cite{TM_07,TM_14} that,
for compact gauge groups
and under the correct relative boundary conditions, one insertion of $Q$ on one
boundary (corresponding to a vertex operator insertion on the boundary
CFT) will, necessarily, demand an insertion of the charge
\be
\bar{Q}=m-\frac{k}{4}n
\lb{ccharge}
\ee
on the other boundary.
We are assuming this fact through the rest of this paper.

Our aim is to orbifold TM theory in a similar way to
Horava~\cite{H_1}, who obtained open boundary world-sheets
through this construction. We are going to take a path integral
approach and reinterpret it in terms of discrete $PT$
and $PCT$ symmetries of the bulk $3D$ TM(GT).

\subsection{Horava Approach to Open World-Sheets\lb{sec:horava}}

Obtaining open string theories out of $3D$ (topological)
gauge theories means building a theory in a manifold which has
boundaries (the $2D$ open string world-sheet) that is already a boundary
(of the $3D$ manifold). This construction raises a problem since
the boundary of a boundary is necessarily a null space.
One interesting way out of this dilemma is to orbifold the $3D$ theory,
then its singular points work as the \textit{boundary} of the
$2D$ boundary.
Horava~\cite{H_1} introduced an orbifold group $G$ that combines the
world-sheet parity symmetry group $Z_2^{WS}$ ($2D$) with two elements
$\{1,\Omega\}$, together with a target symmetry
$\tilde{G}$ of the $3D$ theory fields
\be
G\subset \tilde{G}\times Z_2^{WS}
\ee
With this construction we can get three different kind
of constructions. Elements of the kind $h=\tilde{h}\times 1_{Z_2^{WS}}$
induce twists in the target space (not acting in the world-sheet
at all), for elements $\omega=1_{\tilde{G}}\times\Omega$ we orbifold
the world-sheet manifold (getting an open world-sheet)
without touching in the target space and for elements
$g_1=\tilde{g}_1\times \Omega$ we obtain \textit{exotic} world-sheet orbifold.
In this last case it is further necessary to have an element
corresponding to the twist in the opposite direction
$g_2=\tilde{g_2}\times\Omega$. To specify these twists on
some world-sheet it is necessary to define the 
monodromies of fields on it. Taking the open string
$C_o=C/Z_2$ as the orbifold of the closed string $C$
\be
\pi(C_o)=D\equiv Z_2\ast Z_2\equiv Z_2\subset\hspace{-11.5pt}{\times}Z
\ee
$\ast$ being the free product and $\subset\hspace{-11.5pt}{\times}$
the semidirect product of groups. $D$ is the infinite dihedral group,
the open string first homotopy group. So the monodromies of fields in
$C_o$ corresponds to a representation of this group in the orbifold
group, $Z_2\ast Z_2\to G$, such that the commutative triangle
\be
\ba{lcr}
Z_2\ast Z_2&\longrightarrow&\ \ G\\
\hfill\searrow   &   &\swarrow\hfill\\
           &Z_2^{WS}&
\ea
\ee
is complete. The partition function contains the sum over all possible 
monodromies
\be
Z_C(\tau)=\frac{1}{|G|}\sum_{g_1,g_2,h}Z_C(g_1,g_2,h;\tau)
\ee
where $\tau$ is the moduli of the manifold. The monodromies
$g_1$, $g_2$ and $h$ are elements of $G$ as previously defined
satisfying $g_i^2=1$ and $[g_i,h]=1$.

It will be shown that $PCT$ plays the role of one of such symmetries
with $g_1=g_2$. It is in this sense one of the most simple cases of
\textit{exotic} world-sheet orbifolds.

The string amplitudes can be computed in two different pictures. The
loop-channel corresponds to loops with length $\tau$ of closed and open
strings and the amplitudes are computed as traces over the Hilbert
space. The tree-channel corresponds to a cylinder of length
$\tilde{\tau}$ created from and annihilated to the vacua through
boundary ($\left|B\right>$) and/or crosscaps ($\left|C\right>$) states.
Comparing both ways for the same amplitudes we obtain
\be
\ba{lrcl}
{\rm Annulus\ } (C_2):&\Tr_{\rm open}\left(e^{-H_o \tau}\right)&=&\left<B\right|e^{-H_c\tilde{\tau}}\left|B\right>\vspace{.1 cm}\\
{\rm M\ddot{o}bius\ Strip\ } (M_2):&\Tr_{\rm open}\left(\Omega e^{-H_o \tau}\right)&=&\frac{1}{2}\left<B\right|e^{-H_c\tilde{\tau}}\left|C\right>+\frac{1}{2}\left<C\right|e^{-H_c\tilde{\tau}}\left|B\right>\vspace{.1 cm}\\
{\rm Klein\ Bottle\ } (K_2):\ &\Tr_{\rm open}\left(\Omega e^{-H_o \tau}\right)&=&\left<C\right|e^{-H_c\tilde{\tau}}\left|C\right>
\ea
\lb{tr_BC}
\ee
These equations constitute constraints similar to the modular
invariance constraints of closed string theories. The relation between 
the moduli are, for $K_2$ and $M_2$ $\tau=1/(2\tilde{\tau})$, and
for $C_2$ $\tau=2/(\tilde{\tau})$.

In terms of manifolds it is intended to obtain some open boundary
$\Sigma_{o}=\Sigma/I$ (where boundary refers to
$M=\Sigma\times[0,1]$) which is the involution under the symmetry $I$
of its double, $\Sigma$.
The resulting orbifolded manifold is
\be
M_{o}=(\Sigma\times[0,1])/I
\ee
where $I$ acts in $t$ as Time Inversion $t\to 1-t$.
This construction is presented in figure~\r{fighorava}.
\fig{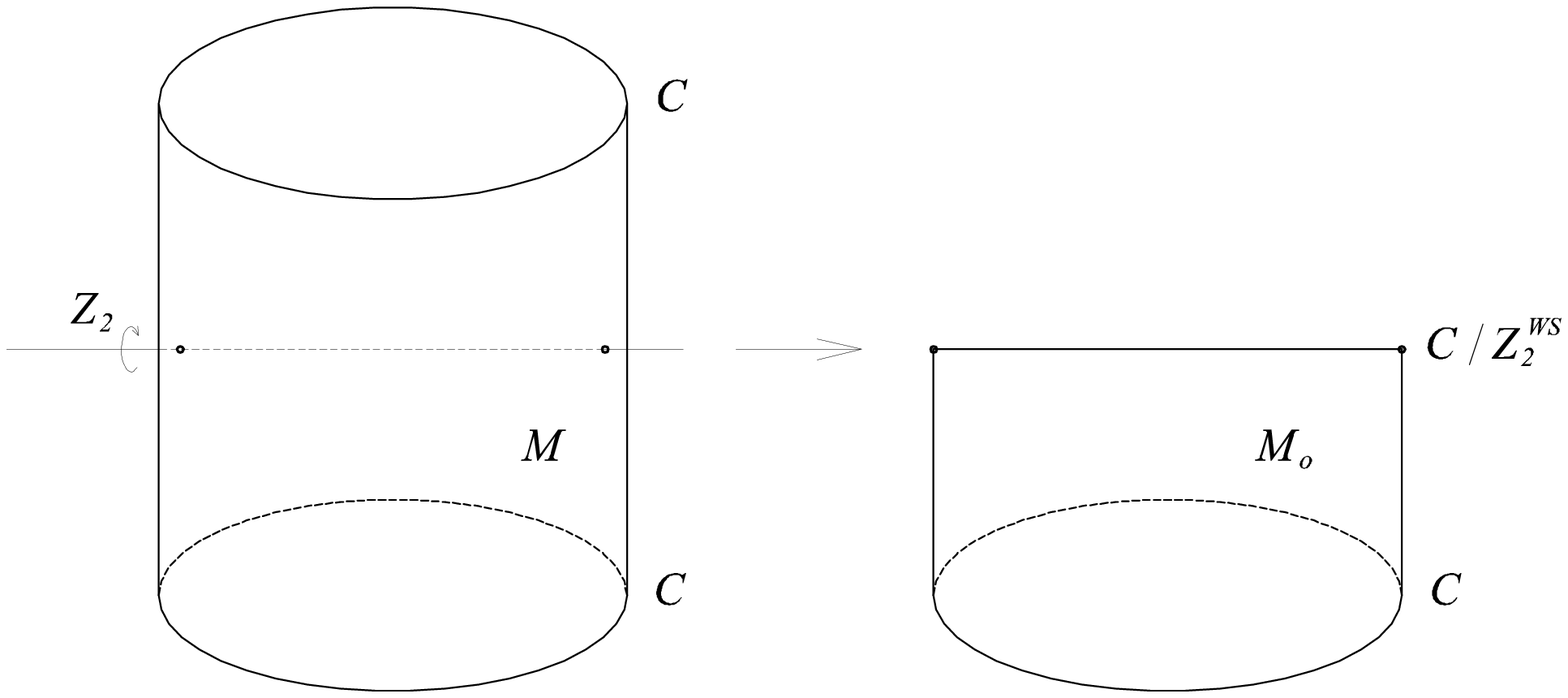}{The thickened open string $C_o$ as an orbifold of the
thickened closed string $C$ under
a $I=Z_2$ symmetry. The boundaries of $C_o=C/Z_2^{WS}$ are the
singular points of the orbifold.}{fighorava}

In terms of the action and fields in the theory Horava used the same
approach of extending them to the doubled manifold
\be
2S_o(A_o)=S(A)
\ee
In simple terms $A$ stands for the extension of $A_o$ from
$M_o$ to its double $M$.

Since there is a one-to-one correspondence between the quantum states
of the gauge theory on M and the blocks of the WZNW model, we may
write
\be
Z_\Sigma=\sum h_{ij}\Psi_i\otimes\bar{\Psi}_j\in {\mathcal H}_\Sigma\otimes\bar{\mathcal H}_\Sigma
\ee
where $\Psi_i$ stands for a basis of the Hilbert space ${\mathcal H}_\Sigma$.
The open string counterpart in the orbifolded theory is
\be
Z_{\Sigma_o}=\sum a_i\Psi_i\in {\mathcal H}_\Sigma
\ee
which also agrees with the fact that in open CFT's the partition
function is the sum of characters (instead of the sum of squares) due
to the holomorphic and antiholomorphic sectors not being independent.

\subsection{Discrete Symmetries and Orbifold of TM(GT)}

Following the discussion of section~\r{sec:riemman} and
section~\r{sec:horava}, it becomes obvious
that the parity operation plays a fundamental role in obtaining
open and/or non-orientable manifolds out of closed orientable ones.
Hence obtaining open/unorientable theories out of closed
orientable theories.

Generally there are several ways of defining parity.
The ones we are interested in have already been presented here.
For the usual ones, $P_1$ and $\Omega$ defined in~(\r{P}) and~(\r{omega}), the fields
of our $3D$ theory transform like
\be
\ba{ccc}
\ba{llcr}
P_1:     &z      &\leftrightarrow&\bz\vspace{.1 cm}\\
         &\Lambda&\rightarrow&\Lambda\vspace{.1 cm}\\
         &A_0    &\rightarrow&A_0\vspace{.1 cm}\\
         &A_z &\leftrightarrow&A_\bz\vspace{.1 cm}\\
         &E^z &\leftrightarrow&E^\bz\vspace{.1 cm}\\
         &B      &\rightarrow&-B\vspace{.1 cm}\\
         &Q      &\rightarrow&Q
\ea&\ \ \ \ &
\ba{llcr}
\Omega:  &z      &\lra       &-\bz\vspace{.1 cm}\\
         &\Lambda&\rightarrow&\Lambda\vspace{.1 cm}\\
         &A_0    &\rightarrow&A_0\vspace{.1 cm}\\
         &A_z    &\lra       &-A_\bz\vspace{.1 cm}\\
         &E^z    &\lra       &-E^\bz\vspace{.1 cm}\\
         &B      &\rightarrow&-B\vspace{.1 cm}\\
         &Q      &\rightarrow&Q
\ea
\ea
\lb{parity}
\ee
where $\Lambda$ is the gauge parameter entering into $U(1)$
gauge transformations. Under these two transformations the action
transforms as
\be
\int(F^2+kA\wedge F)\rightarrow\int(F^2-kA\wedge F)
\lb{PTS}
\ee

The theory is clearly not parity invariant. Let us then look for further
discrete symmetries which we may combine with parity in order to make
the action (theory) invariant. Introduce time-inversion, $T: t \to 1-t$,
implemented in this non-standard way due to the compactness of time.
Note that $t=1/2$ is a fixed point of this operation.
Upon identification of the boundaries as described in~\cite{TM_14}
the boundary becomes a fixed point as well. It remains to define
how the fields of the theory change under this symmetry.
There are two possible transformations compatible with gauge transformations,
$A_\Lambda(t,z,\bz)=A(t,z,\bz)+\partial\Lambda(t,z,\bz)$. They are:
\be
\ba{cccr}
CT:&t      &\rightarrow&1-t\vspace{.1 cm}\\
    &\Lambda&\rightarrow&\Lambda\vspace{.1 cm}\\
    &A_0    &\rightarrow&-A_0\vspace{.1 cm}\\
    &\vb{A} &\rightarrow&\vb{A}\vspace{.1 cm}\\
    &\vb{E} &\rightarrow&-\vb{E}\vspace{.1 cm}\\
    &B      &\rightarrow&B\vspace{.1 cm}\\
    &Q      &\rightarrow&Q
\ea
\lb{T1}
\ee
and
\be
\ba{cccr}
T:&t      &\rightarrow&1-t\vspace{.1 cm}\\
    &\Lambda&\rightarrow&-\Lambda\vspace{.1 cm}\\
    &A_0    &\rightarrow&A_0\vspace{.1 cm}\\
    &\vb{A} &\rightarrow&-\vb{A}\vspace{.1 cm}\\
    &\vb{E} &\rightarrow&\vb{E}\vspace{.1 cm}\\
    &B      &\rightarrow&-B\vspace{.1 cm}\\
    &Q      &\rightarrow&-Q
\ea
\lb{T2}
\ee
where we defined $C$, charge conjugation, as $A_\mu\to-A_\mu$.
This symmetry inverts the sign of the charge, $Q\to-Q$, as usual.
These discrete symmetries together with parity $P$ or $\Omega$ are the
common ones used in $3D$ Quantum Field Theory.
When referring to parity in generic terms we will use the letter $P$.

Under any of the $T$ and $CT$ symmetries the action changes
in the same fashion it does for parity $P$, as given by~(\r{PTS}).
In this way any of the combinations $PT$ and $PCT$ are symmetries
of the action, $S\rightarrow S$.
Gauging them is a promising approach to define the TM(GT) orbifolding.
It is now clear why we need extra symmetries,
besides parity, in order to have combinations of them under which 
the theory (action) is invariant.
In general, whatever parity definition we use, these results
imply that $PT$ and $PCT$ are indeed symmetries of the theory.

We can conclude straight away that any of the two previous symmetries
exchange \textit{physically} two boundaries working as a mirror
transformation with fixed \textit{point} \mbox{$(t=1/2,z=\bz=x)$}
(corresponds actually to a line) as pictured in figure~\r{fig.mirror}.
We are considering that, whenever there is a charge insertion in
one boundary of $\mbox{q=m+kn/4}$, it will exist an
insertion of $\bar{q}=m-kn/4$ in the other
boundary~\cite{TM_05,TM_14}.

Under the symmetries $PT$ and $PCT$ as given by~(\r{parity}), (\r{T1})
and~(\r{T2}) the boundaries will be exchanged as presented in
figure~\r{fig.mirror}. In the case of $PCT$ the charges will simply be 
swaped but in the case of $PT$ their sign will be change $q\to-q$.
Note that $\Sigma_{\frac{1}{2}}=\Sigma(t=1/2)$ only feels $P$ or $CP$.
\fig{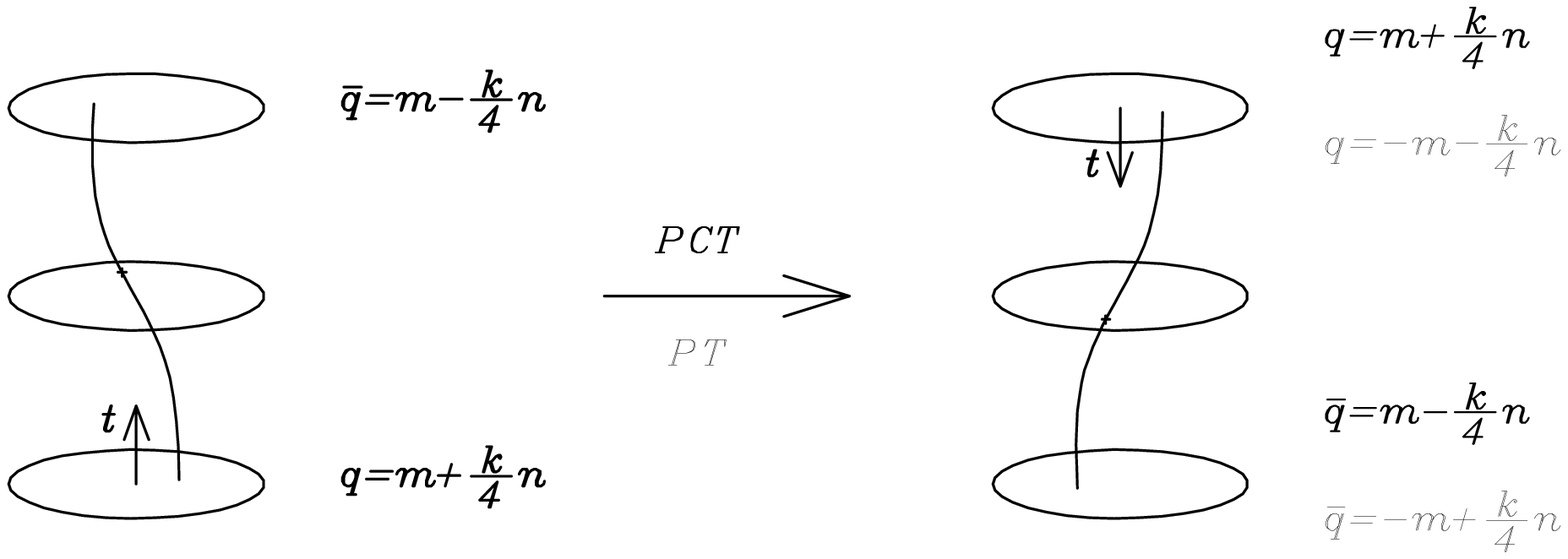}{Exchange of boundaries due to $PT$/$PCT$ transformation.}{fig.mirror}

As will be shown in detail there are important differences between the
two symmetries $CT$ and $T$, they will effectively gauge field
configurations corresponding to untwisted/twisted sectors of closed
strings and Neumann/Dirichlet boundary conditions of open strings.

Not forgetting that our final aim is to orbifold/quotient our theory
by gauging the discrete symmetries, let us proceed to
check compatibility with the desired symmetries in detail.
It is important to stress that field configurations satisfying any $PT$/$PCT$
combinations of the previous symmetries exist, in principle,
from the start in the theory. We can either impose by hand that the
physical fields obey one of them (as is usual in QFT)
or we can assume that we have a wide theory with all of these field
configurations
and obtain (self consistent) \textit{subtheories} by
building suitable projection operators that select some type of configurations.
It is precisely this last construction that we have in mind when building
several different theories out of one. In other words we are going
to build different \textit{new} theories by gauging discrete
symmetries of the type $PCT$ and $PT$. 

It is important to stress what the orbifold means in terms of the
boundaries and bulk from the point of view of TM(GT).
It is splitting the manifold $M$ into two pieces creating one new
boundary at $t=1/2$. This boundary is going to feel only $CP$ or $P$
symmetries since it is located at the temporal fixed point of the
orbifold. Figure~\r{fig.orbi} shows this procedure.
\fig{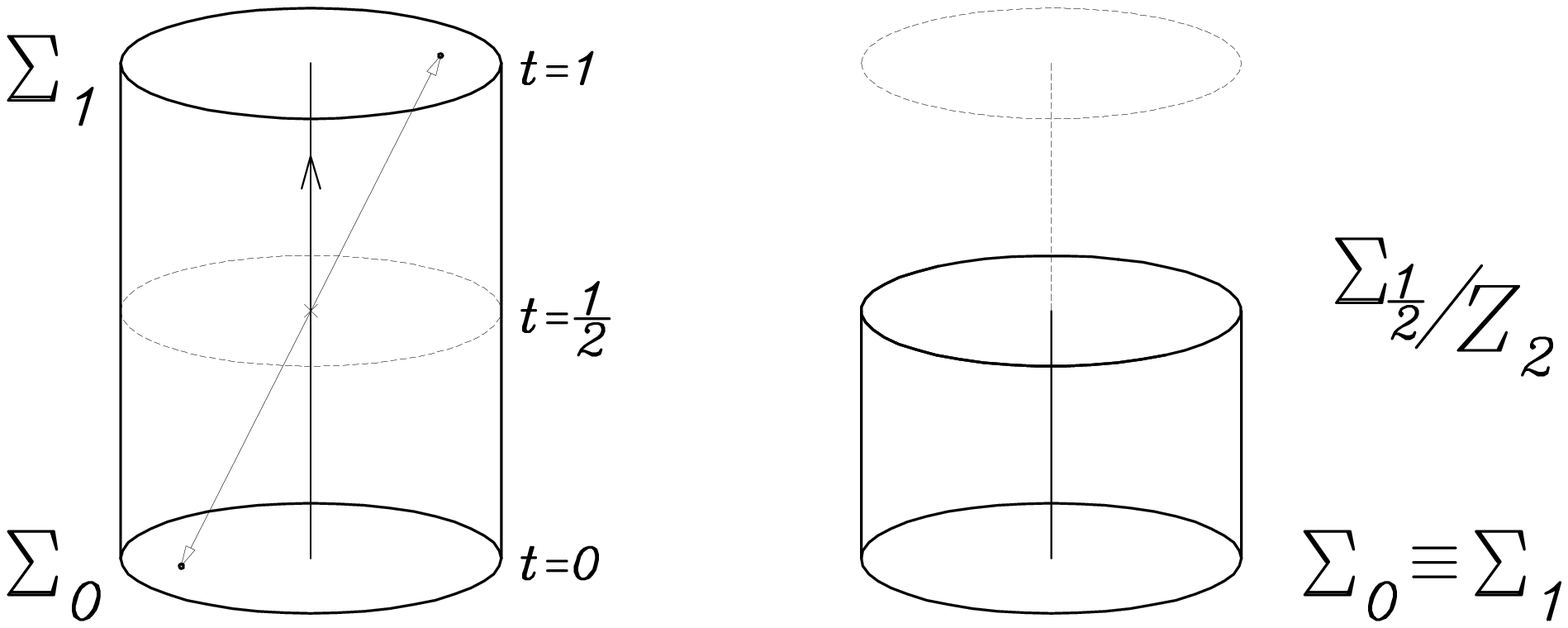}{Orbifolding of TM(GT). $\Sigma_{\frac{1}{2}}$ only
feels $PT$ or $P$ which are isomorphic to $Z_2$}{fig.orbi}
In this way this new boundary is going to constrain the \textit{new}
theory in such a way that the boundary theories will correspond 
to open and unoriented versions of the original full theory.

\subsection{Tree Level Amplitudes for\\ Open and Closed Unoriented Strings}

We start by considering tree level approximation to string amplitudes,
i.e. the Riemann surfaces are of genus $0$.  These surfaces are
the sphere (closed oriented strings) and its orbifolds: the disk (open
oriented) and the projective plane (closed unoriented) as was
discussed in section~\r{sec:riemman}.  From the point of view of
TM(GT), orbifolding means that we split the manifold $M$ into two
pieces that are identified.  As a result at $t=1/2$, the fixed point
of the orbifold, a new boundary is created.

For different orbifolds we shall have different admissible field
configurations. In the following discussion we studied which are the
configurations compatible with $PT$ and $PCT$ for the several parity
operations already introduced.

\subsubsection{Disk}

Let us start from the simplest case - the disk is obtained by
the involution of the sphere under $P_1$ as given
by~(\r{P}). So consider the identifications under $P_1CT$ and $P_1T$.
For the first one the fields relate as
\be
\ba{rrcl}
P_1CT:&\Lambda(t,z,\bz)      &=&\Lambda(1-t,\bz,z)\vspace{.1 cm}\\
            &\partial_iE^i(t,z,\bz)&=&-\partial_iE^i(1-t,\bz,z)\vspace{.1 cm}\\
            &B(t,z,\bz)            &=&-B(1-t,\bz,z)\vspace{.1 cm}\\
            &\displaystyle Q(t)=\int_{\Sigma(t)}\rho_0 &=&\displaystyle Q(1-t)=-\int_{\tilde{\Sigma}(1-t)}(-\rho_0)
\ea
\lb{PCT}
\ee
The orientations of
$\Sigma$ and $\tilde{\Sigma}$ are opposite.
Under these relations the Wilson lines have the property
\be
\exp{\left\{iQ\int_C dx^\mu A_\nu\right\}}=\exp{\left\{iQ\int_{-C} dx^\mu A_\nu\right\}}
\lb{PTW}
\ee
This means that for the configurations obeying the
relations~(\r{PCT}) we loose the notion of time direction.

Under the \textit{involution} of our $3D$ manifold,
using the above relations as geometrical identifications, the boundary
becomes $t=0$ and $t=1/2$.
For the moment let us check the compatibility of the
observables with the proposed orbifold constructions given by the
previous relations. In a very naive and straightforward way, when
we use $P_1CT$ as given by (\r{PCT}) the
charges should maintain their sign ($q(t)\cong q(1-t)$). Then by exchanging
boundaries we need to truncate the spectrum and set $q\cong{\bar{q}}=m$
in order the identification to make sense.
Let us check what happens at the singular point of our orbifolded
theory, $t=1/2$. The fields are identified according to the previous
rules but the manifold $\Sigma(t=1/2)=S^2$ is only affected by $P_1$.

Take two Wilson lines that pierce the
manifold in two distinct points, $z$ and $z'$. Under the previous
involution $P_1CT$, $z$ is identified with $\bz$ for $t=1/2$.
Then, geometrically, we must have $z'=\bz$ in order to have spatial
identification of the piercings. The problem is that
when we have only two Wilson lines, TM(GT) demands that they carry
opposite charges. In order to implement the desired identification
we are left with $q=0$ as the only possibility.
For the case where the Wilson lines pierce the manifold in the real
axis, $z=x$ and $z'=x'$, the involution is possible
as pictured in figure~\r{fig.s2pct1a} since we identify
$x\cong x$ and $x'\cong x'$.

In the presence of three Wilson lines,
following the same line of arguing, we
will necessarily have one insertion in the boundary and two in the bulk
as pictured in figure~\r{fig.s2pct1b}. Only in the presence of
four Wilson lines, as pictured in figure~\r{fig.s2pct1c} can we
avoid any insertion in the boundary.

\fig{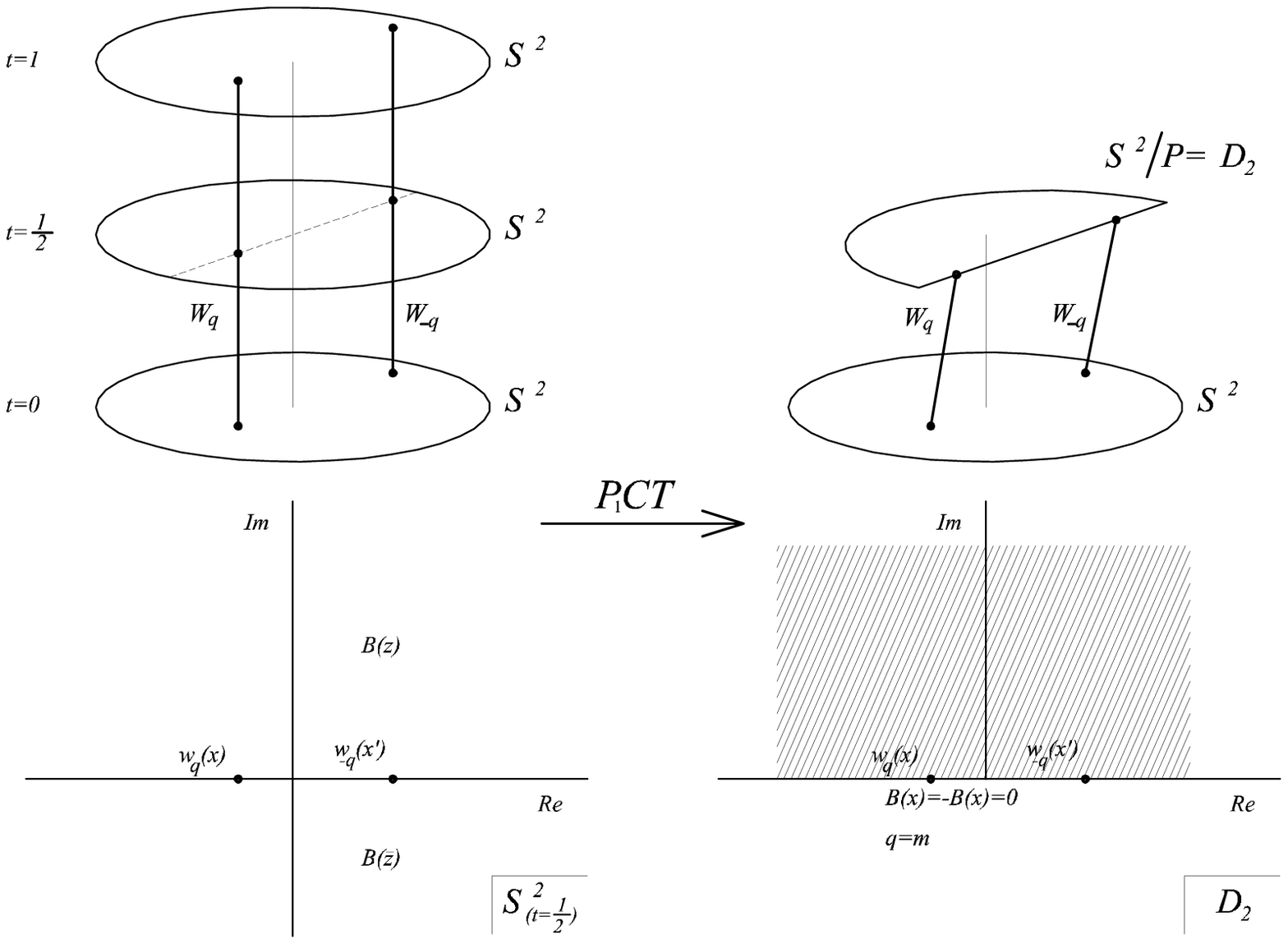}{Orbifold under $P_1CT$ in the presence of 2 Wilson
lines, $W_q$ and $W_{-q}$. They need to pierce
$\Sigma_{o\frac{1}{2}}=S^2/P_1=D_2$ in the real axis and the allowed
charges are $q=m$.}{fig.s2pct1a}\vspace{-.3cm}
\fig{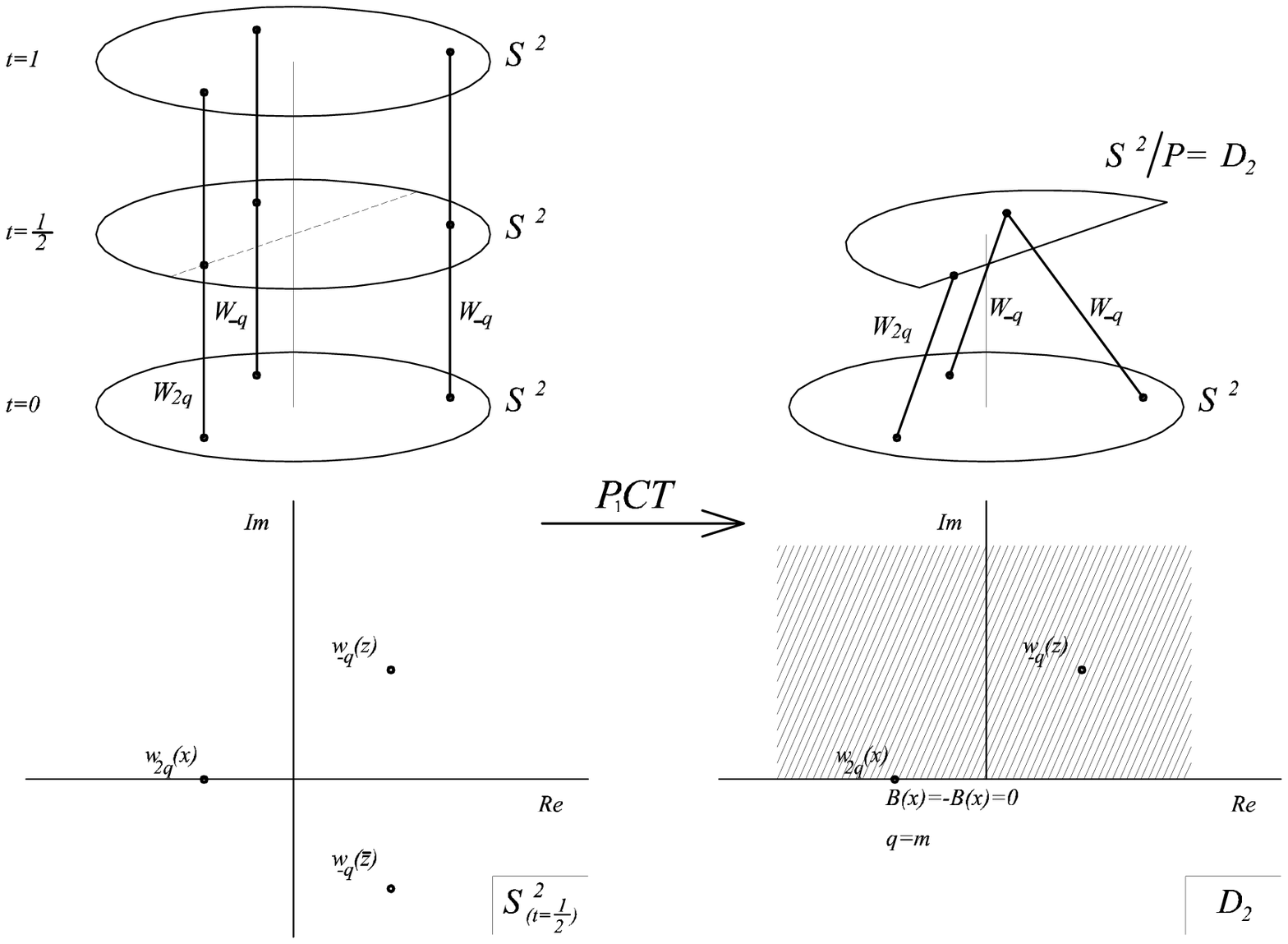}{Orbifold under $P_1CT$ in the presence of 3 Wilson
lines, $W_{2q}$ and two $W_{-q}$. $W_{2q}$ must pierce
$\Sigma_{o\frac{1}{2}}=S^2/P_1=D_2$ in the real axis and the allowed
charges are $q=m$.}{fig.s2pct1b}
\clearpage

\fig{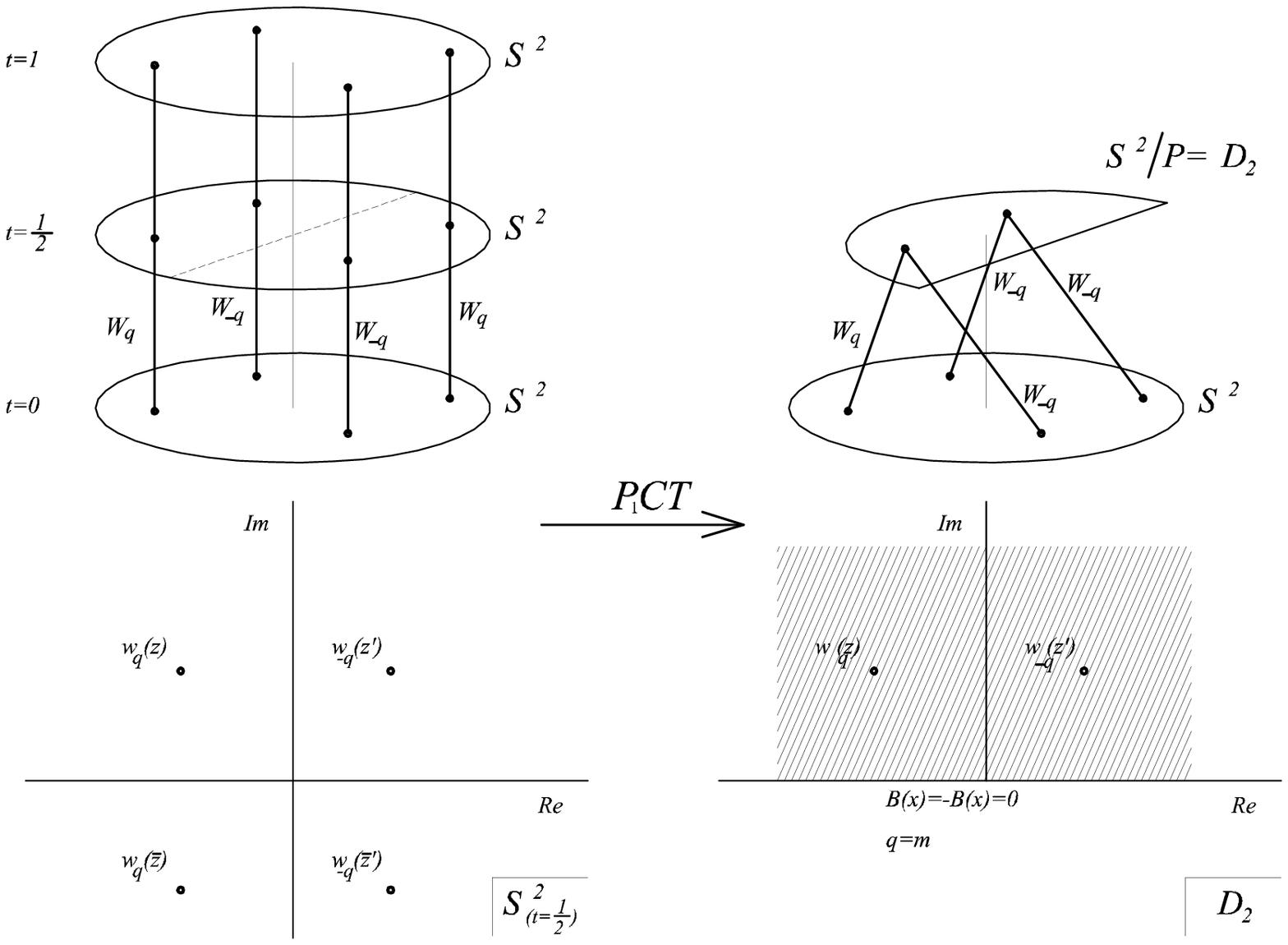}{Orbifold under $P_1CT$ in the presence of 4 Wilson
lines, two $W_q$ and two $W_{-q}$. The allowed
charges are $q=m$.}{fig.s2pct1c}

Note that the identification
$B(z,\bz)\cong-B(\bz,z)$ in the real axis implies necessarily
$B(x,x)=0$. Remember that $2\pi n=\int B$ (see~\cite{TM_07,TM_14} for details).
We could as well have an insertion in the boundary and one in the bulk

This fact is simply the statement that by imposing $P_1CT$ we are
actually imposing {\bf Neumann} boundary conditions. The charges
of the theory become $q=m$, this means that the string spectrum
has only Kaluza-Klein momenta. Furthermore the monopole induced processes
are suppressed, recall that they change the charge by an amount
$kn/2$ which would take the charges out of the spectrum allowed
in this configurations.

Following our journey consider next $P_1T$. The fields now are related
in the following way
\be
\ba{rrcl}
P_1T:&\Lambda(t,z,\bz)       &=&-\Lambda(1-t,\bz,z)\vspace{.1 cm}\\
            &\partial_iE^i(t,z,\bz) &=&\partial_iE^i(1-t,\bz,z)\vspace{.1 cm}\\
            &B(t,z,\bz)             &=&B(1-t,\bz,z)\vspace{.1 cm}\\
            &\displaystyle Q=\int_{\Sigma(t)}\rho_0&=&-Q=-\int_{\tilde{\Sigma}(1-t)}(\rho_0)
\ea
\lb{PT}
\ee
The Wilson line has the same property~(\r{PTW}) as in the previous case.

Now the charges change sign under a $P_1T$ symmetry.
As before identifying the charges in opposite boundaries
truncates the spectrum, $q(t)\cong-q(1-t)$.
So we must have $q\cong-{\bar{q}}=nk/4$.

We can, in this case identify two piercings in the bulk since the charge
identifications are now $q\cong-q$ is compatible with TM(GT).
But we cannot insert any operator other than the identity $\phi_0$ in the
real axis since the corresponding charge must be zero $q(x)=-q(x)=0$.
Therefore this kind of orbifolding is only possible when we have a
even number of Wilson lines propagating in the bulk.
The result for two Wilson lines is pictured in figure~\r{fig.s2idpt1a}
and for four in figure~\r{fig.s2idpt1b}.
\figh{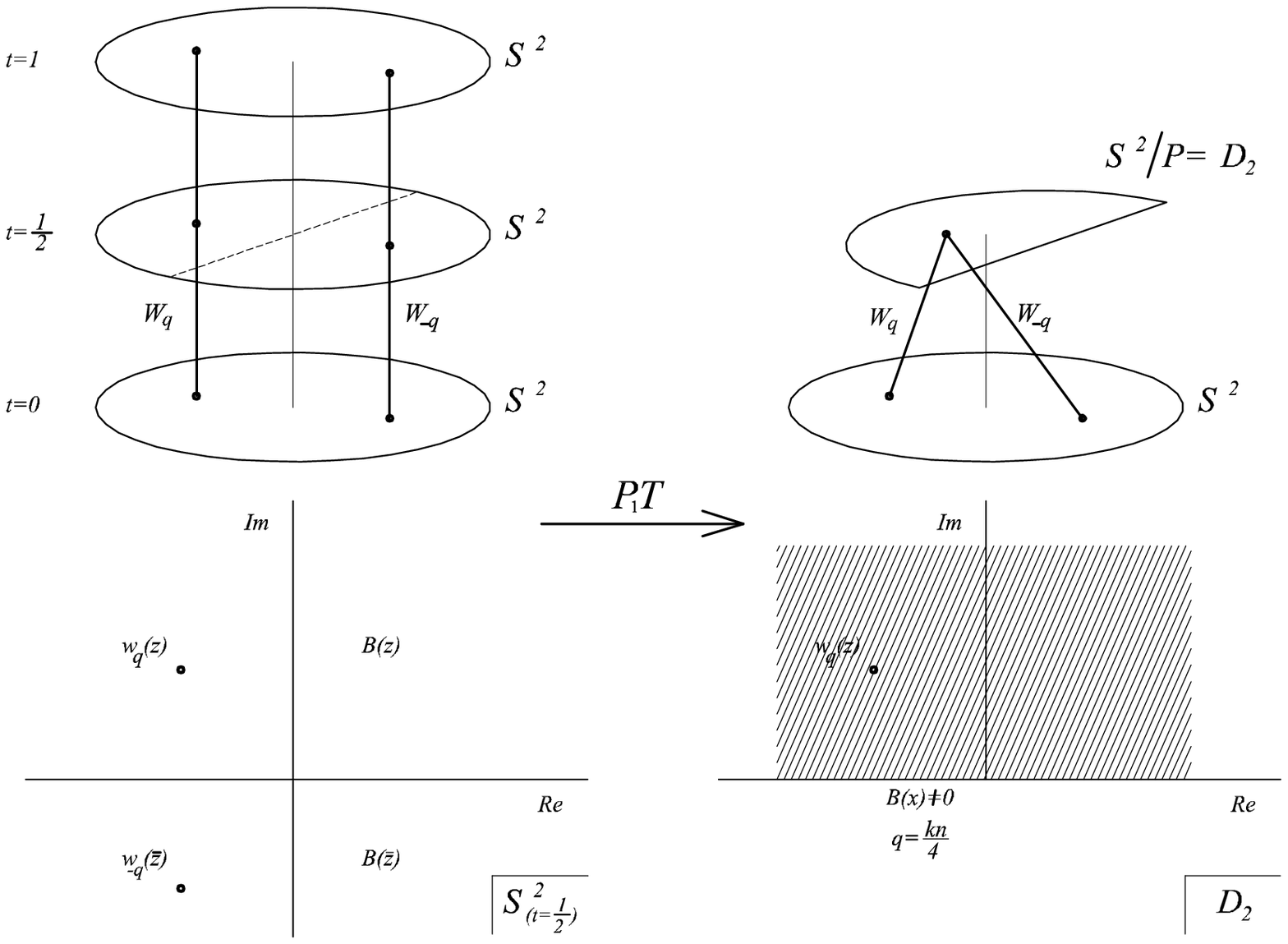}{Orbifold under $P_1T$ in the presence of 2 Wilson
lines, $W_q$ and $W_{-q}$. The new boundary is
$\Sigma_{o\frac{1}{2}}=S^2/P_1=D_2$. The allowed
charges are $q=kn/4$.}{fig.s2idpt1a}

\fig{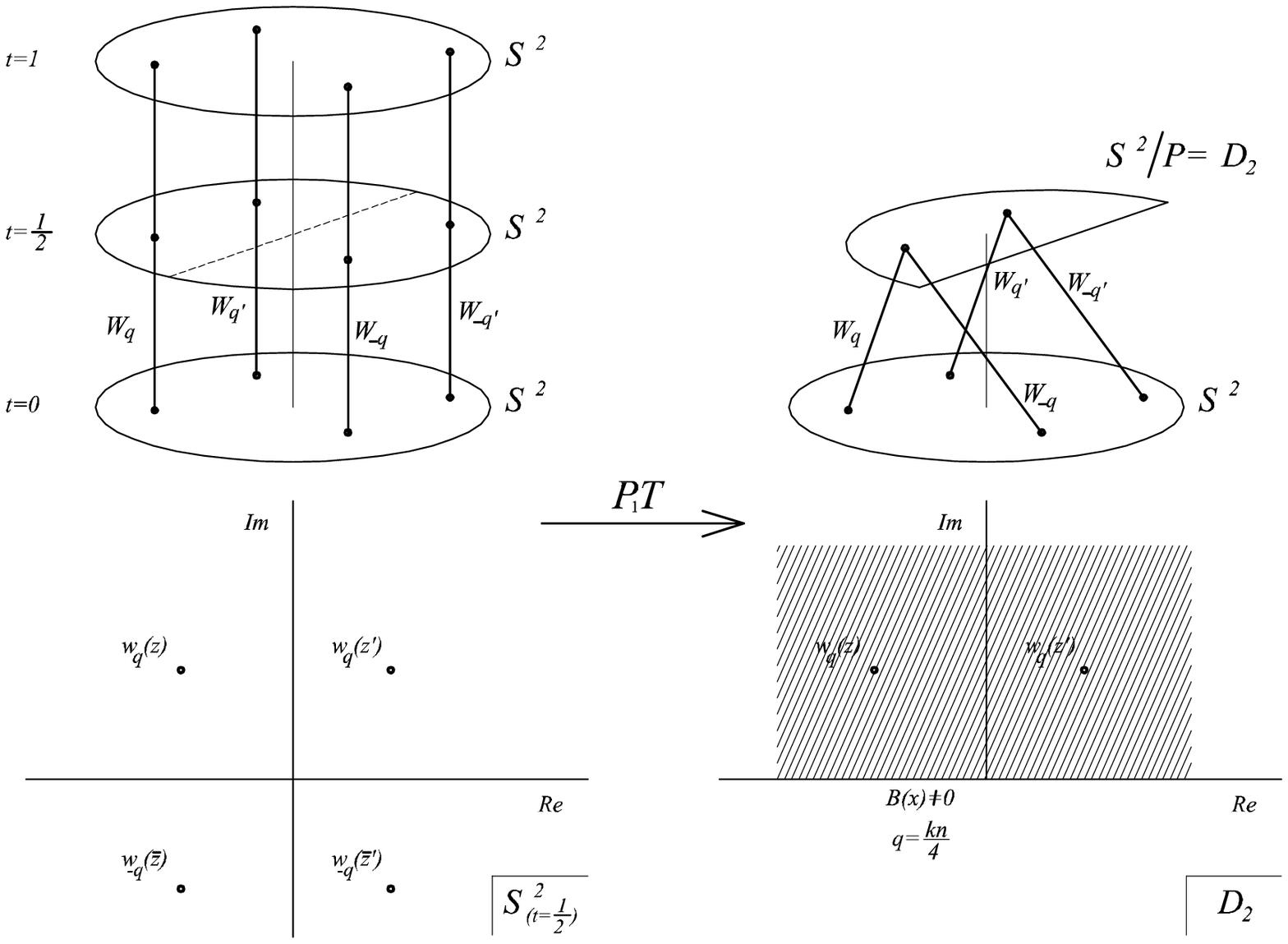}{Orbifold under $P_1T$ in the presence of 4 Wilson
lines, $W_q$, $W_q'$, $W_{-q}$ and $W_{-q'}$. The new boundary is
$\Sigma_{o\frac{1}{2}}=S^2/P_1=D_2$. The allowed
charges are $q=kn/4$.}{fig.s2idpt1b}

In terms of the full theory, we just define a new $2D$ boundary which
is a disk. The piercings of Wilson lines are none others than vertex operators
(or fields) of a Conformal Field Theory defined on the Disk. 
In this case $B(z,\bz)=B(\bz,z)$, then $B\neq 0$ in the boundary.

So this facts translates into {\bf Dirichlet}
boundary conditions for $P_1T$. The charges allowed are $q=nk/4$,
the winding number of string theory. The monopole induced processes
are now allowed being crucial in this construction since they allow
the \textit{gluing} in this new boundary of two Wilson lines carrying
charges $q=nk/4$ and $\bar{q}=-nk/4$.

\subsubsection{Projective Plane}

We now consider the parity operation as the antipodal identification 
given in~(\r{PP}). We thus obtain the projective plane as the \textit{new} $2d$
boundary of TM(GT).

The transformation is given by
\be
\ba{rrcl}
P_2: &z           &\rightarrow &\displaystyle-\frac{1}{\bz}\vspace{.1 cm}\\
   &\bz         &\rightarrow &\displaystyle-\frac{1}{z}\vspace{.1 cm}\\
   &\Lambda(z,\bz)  &\rightarrow &\displaystyle\Lambda(-\frac{1}{\bz},-\frac{1}{z})\vspace{.1 cm}\\
   &A_z(z,\bz)  &\rightarrow &\displaystyle\frac{1}{z^2} A_\bz(-\frac{1}{\bz},-\frac{1}{z})\vspace{.1 cm}\\
   &A_\bz(z,\bz)&\rightarrow &\displaystyle\frac{1}{\bz^2} A_z(-\frac{1}{\bz},-\frac{1}{z})\vspace{.1 cm}\\
   &B(z,\bz)&\rightarrow &\displaystyle\frac{1}{z^2\bz^2} B(-\frac{1}{\bz},-\frac{1}{z})\vspace{.1 cm}\\
   &E_z(z,\bz)&\rightarrow &\displaystyle\frac{1}{z^2} E_\bz(-\frac{1}{\bz},-\frac{1}{z})\vspace{.1 cm}\\
   &E_\bz(z,\bz)&\rightarrow &\displaystyle\frac{1}{\bz^2} E_z(-\frac{1}{\bz},-\frac{1}{z})\vspace{.1 cm}\\
\ea
\ee
Again we proceed to check the compatibility of the identifications
under this new discrete symmetry, that is $t'\cong 1-t$,
$z'\cong-1/\bz$ and $\bz'\cong-1/z$. We obtain for $P_2CT$
\be
\ba{rrcl}
P_2CT:&\Lambda(t,z,\bz)      &=&\Lambda(1-t,-\frac{1}{\bz},-\frac{1}{z})\vspace{.1 cm}\\
     &\partial_iE^i(t,z,\bz)&=&-\frac{1}{z^2\bz^2}\partial_iE^i(1-t,-\frac{1}{\bz},-\frac{1}{z})\vspace{.1 cm}\\
     &B(t,z,\bz)            &=&-\frac{1}{z^2\bz^2}B(1-t,-\frac{1}{\bz},-\frac{1}{z})\vspace{.1 cm}\\
     &\displaystyle Q(t)=\int_{\Sigma(t)}\rho_0 &=&\displaystyle Q(1-t)=-\int_{\tilde{\Sigma}(1-t)}(-\rho_0)
\ea
\lb{PPCT}
\ee
Note that the relation between the integrals
\be
\int_{\tilde{\Sigma}(t')}\frac{d^2z'}{z'^2\bz'^2}z'^2\bz'^2\left(B(t',z',\bz')+\partial_iE^i(t',z',\bz')\right)=\int_{\Sigma(t)}d^2z\left(B(t,z,\bz)+\partial_iE^i(t,z,\bz)\right)
\lb{Qzz}
\ee
follows from taking into account the second and third equalities of~(\r{PPCT}),
and the relations $dz=d\bz'/\bz'^2$, $d\bz=dz'/z'^2$, and consequently
$dz\wedge d\bz=-(1/z'\bz')dz'\wedge d\bz'$.
$\Sigma$ and $\tilde{\Sigma}$ again have opposite orientations
and are mapped into each other by the referred involution.
Under these relations and in a similar way to~(\r{Qzz})
the action transforms under $P_2$ as given in (\r{PTS}) and any of the
combinations $P_2CT$ or $P_2T$ keep it invariant.
Also the Wilson lines have the same property given by (\r{PTW}).

In the derivation of the previous identifications
(\r{PPCT}) we had to demand analyticity of the fields on the full
sphere. This translates into demanding the transformation between the
two charts covering the sphere to be well defined.
Since $\partial_u\Lambda=-z^2\partial_z\Lambda$  
and $\partial_{\bar{u}}\Lambda=-\bz^2\partial_\bz\Lambda$ the fields
must behave at infinity and zero like
\be
\ba{cc}
\Lambda\stackrel{\infty}{\to} z^{-1}\bz^{-1}&\Lambda\stackrel{0}{\to} z^3\bz^3\vspace{.1 cm}\\
A_z\stackrel{\infty}{\to} z^{-2}\bz^{-1}&A_z\stackrel{0}{\to} z^2\bz\vspace{.1 cm}\\
A_\bz\stackrel{\infty}{\to} z^{-1}\bz^{-2}&A_\bz\stackrel{0}{\to} z\bz^2
\ea
\ee

If naively we didn't care about these last limits the relations would
be plagued with Dirac deltas coming
from the identity $2\pi\delta^2(z,\bz)=\partial_z(1/\bz)=\partial_\bz(1/z)$.
Once the previous behaviors are taken into account all these terms will
vanish upon integration. Another way to interpret these results is to
note that the points at infinity are not part of the chart (not
physically meaningful), to check the physical behavior at those
points we have to compute it at zero in the other chart.

This time the charges compatible with $P_2CT$ are $q=m$ since
$q\cong\bar{q}$. Once there are no boundaries it is not possible
to have configurations with two Wilson which allow this kind of orbifold.
In this way the lowest number of lines is four as pictured in
figure~\r{fig.s2idpct2}. Furthermore the number of Wilson lines must be
even.
\fig{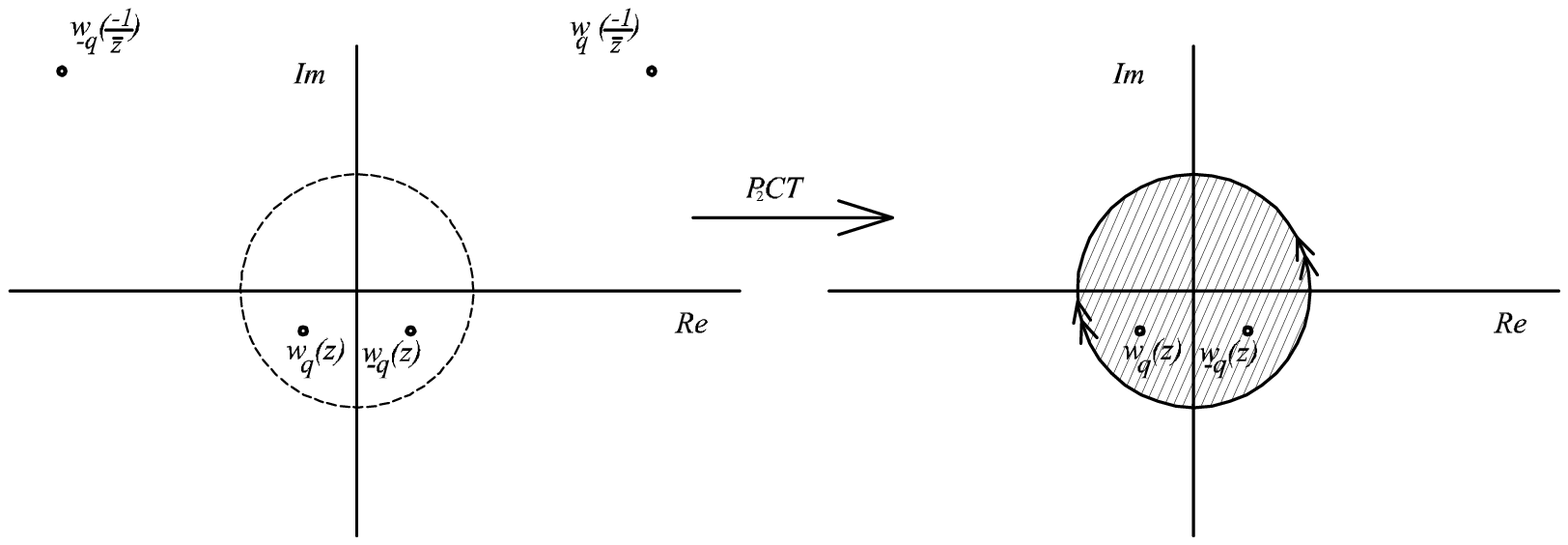}{Orbifold under $PCT$ in the presence of 4 Wilson
lines, two $W_q$ and two $W_{-q}$. The new $2d$ boundary is
$\Sigma_{o\frac{1}{2}}=S^2/P=RP_2$. The allowed
charges are $q=m$.}{fig.s2idpct2}

This configuration corresponds to {\bf untwisted} closed unoriented string
theories. Note that $\Lambda$, which is identified with string theory
target space, is not orbifolded by $P_2CT$. The charges allowed are $q=m$,
the KK momenta of string theory. Once again the monopole processes
are suppressed.

For $P_2T$ the fields relate as
\be
\ba{cccc}
P_2T:&\Lambda(t,z,\bz)      &=&-\Lambda(1-t,-\frac{1}{\bz},-\frac{1}{z})\vspace{.1 cm}\\
     &\partial_iE^i(t,z,\bz)&=&\frac{1}{z^2\bz^2}\partial_iE^i(1-t,-\frac{1}{\bz},-\frac{1}{z})\vspace{.1 cm}\\
     &B(t,z,\bz)            &=&\frac{1}{z^2\bz^2}B(1-t,-\frac{1}{\bz},-\frac{1}{z})\vspace{.1 cm}\\
     &\displaystyle Q(t)=\int_{\Sigma(t)}\rho_0 &=&\displaystyle-Q(1-t)=-\int_{\tilde{\Sigma}(1-t)}(-\rho_0)
\ea
\lb{PPT}
\ee

In this case $q=kn/4$ since $q\cong-\bar{q}$ and further
configurations with two Wilson lines are compatible with the orbifold
as pictured in figure~\r{fig.s2idpt2}.
\figh{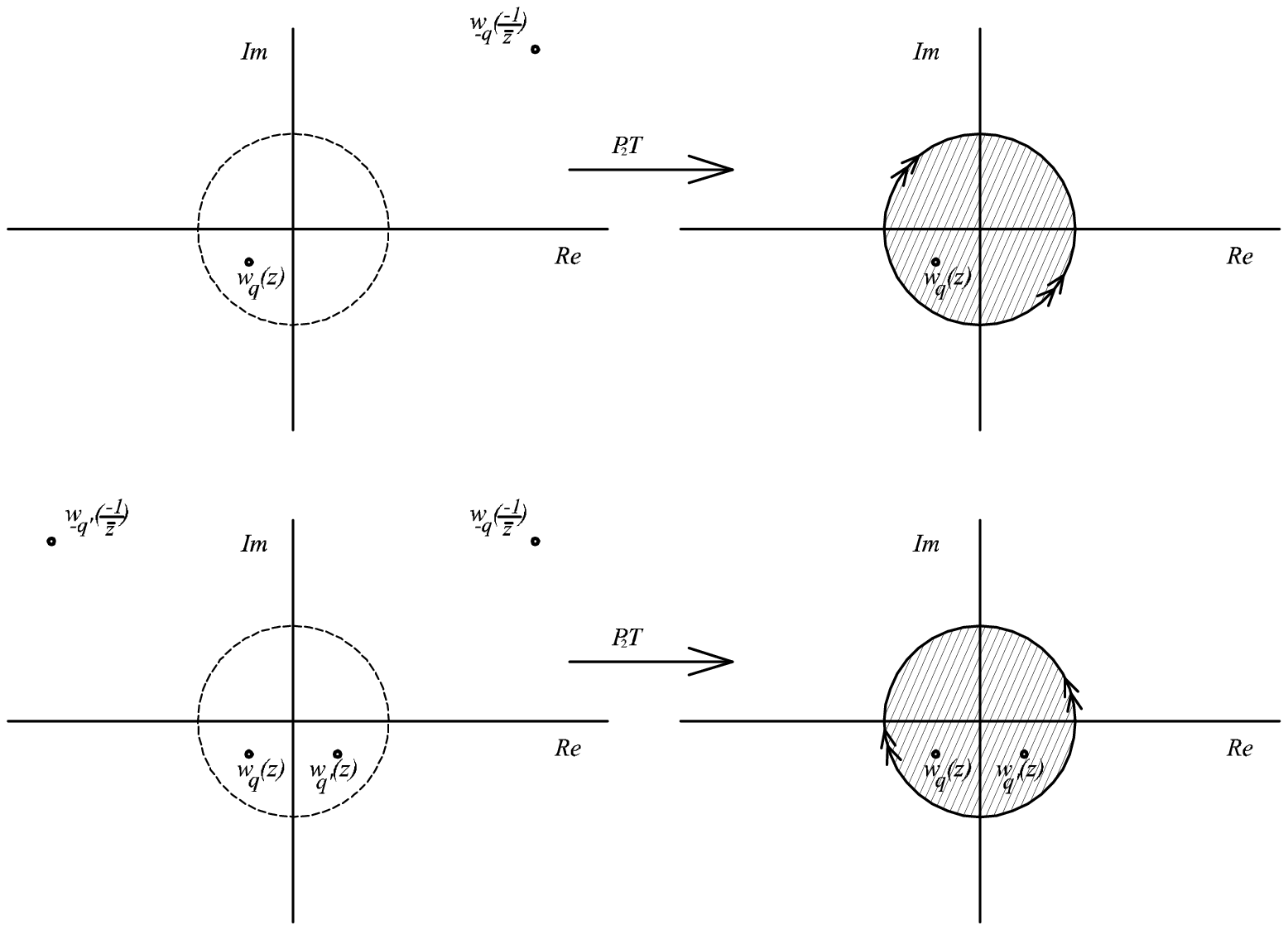}{Orbifold under $PCT$ in the presence of 4 Wilson
lines, two $W_q$ and two $W_{-q}$. The new $2d$ boundary is
$\Sigma_{o\frac{1}{2}}=S^2/P=RP_2$. The allowed
charges are $q=m$.}{fig.s2idpt2}

In this case we have {\bf twisted} unoriented closed strings. Note
that the orbifold identifies $\Lambda\cong-\Lambda$ such that the
target space of string theory is orbifolded. The full construction,
including the world-sheet parity,
from the point of view of string theory is called an orientifold.
The allowed charges $q=kn/4$ correspond to the winding number of
string theory. The monopole processes are again
crucial since allow, in the new boundary, the
\textit{gluing} of Wilson lines carrying opposite charges. We will
return to this discussion.

\subsection{One Loop Amplitudes for\\ Open and Closed Unoriented Strings}

One loop amplitudes are computed for Riemann surfaces of genus $1$.
They correspond to the torus (closed oriented) and its orbifolds:
the annulus or cylinder (open oriented), the M\"{o}bius strip
(open unoriented) and the Klein bottle (closed unoriented).

\subsubsection{Annulus}
We start with the already
studied parity transformation $\Omega$, as given by (\r{omega}).
There is nothing new to add to the fields relations (\r{PCT}) for
$PCT$ and (\r{PT}) for $PT$, this time under the
identifications $t'=1-t$, $z'=-\bz$ and $\bz'=-z$. The resulting geometry
is the annulus $C_2$ and has now two boundaries.

\figh{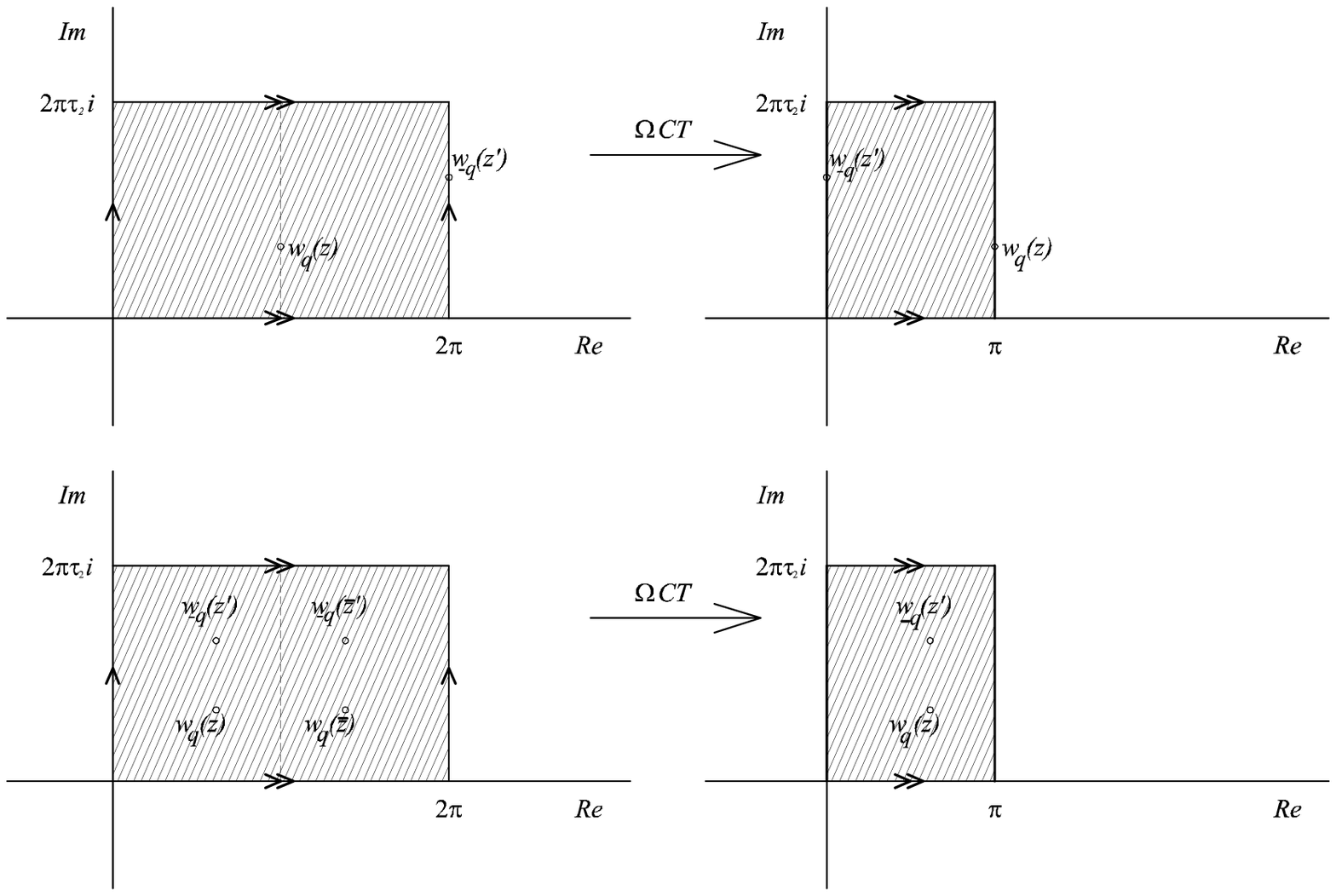}{Orbifold under $\Omega CT$ in the presence of 2 and 4 Wilson
lines. The new $2d$ boundary is
$\Sigma_{o\frac{1}{2}}=S^2/\Omega=C_2$. The allowed
charges are $q=m$.}{fig.t2idpct1}

For $\Omega CT$ the allowed charges are $q=m$ due to the identification
$q\cong\bar{q}$ and $B(x)=0$ at the boundaries.
We can have two insertions in the boundaries of the
$2d$ CFT but not in the bulk due to the identifications of charges,
basically the argument is the same as used for the disk.
As in the disk we cannot have one single bulk insertion due
to the total charge being necessarily zero in the full plane.
Up to configurations with four Wilson lines we can have:
two insertions in the boundary; one insertion in the
bulk and one in the boundary corresponding to three Wilson lines;
three insertions in the boundaries (with $\sum q=0$); one insertion
in the bulk and two in the boundary corresponding to four Wilson
lines; and two insertions in the bulk corresponding to four Wilson
lines as pictured in figure~\r{fig.t2idpct1}.
This construction corresponds to open oriented strings with
{\bf Neumann} boundary conditions. The charge spectrum is $q=m$,
corresponding to KK momenta in string theory and the monopole
induced processes are suppressed. It is Neumann because the gauged
symmetry is of $PCT$ type. We note that the definition of parity is
not important, even for genus 1 surfaces the results hold similarly to 
the previous cases for $P_1$ and $P_2$ used in genus 0. What is
important is the inclusion of the discrete symmetry $C$!

For $\Omega T$ the allowed charges are $q=kn/4$
due to the identification $q\cong-\bar{q}$. There are no insertions in
the boundary. One insertion in the bulk corresponds to two Wilson
lines and two to four Wilson lines presented in picture~\r{fig.t2idpt1}.
\figh{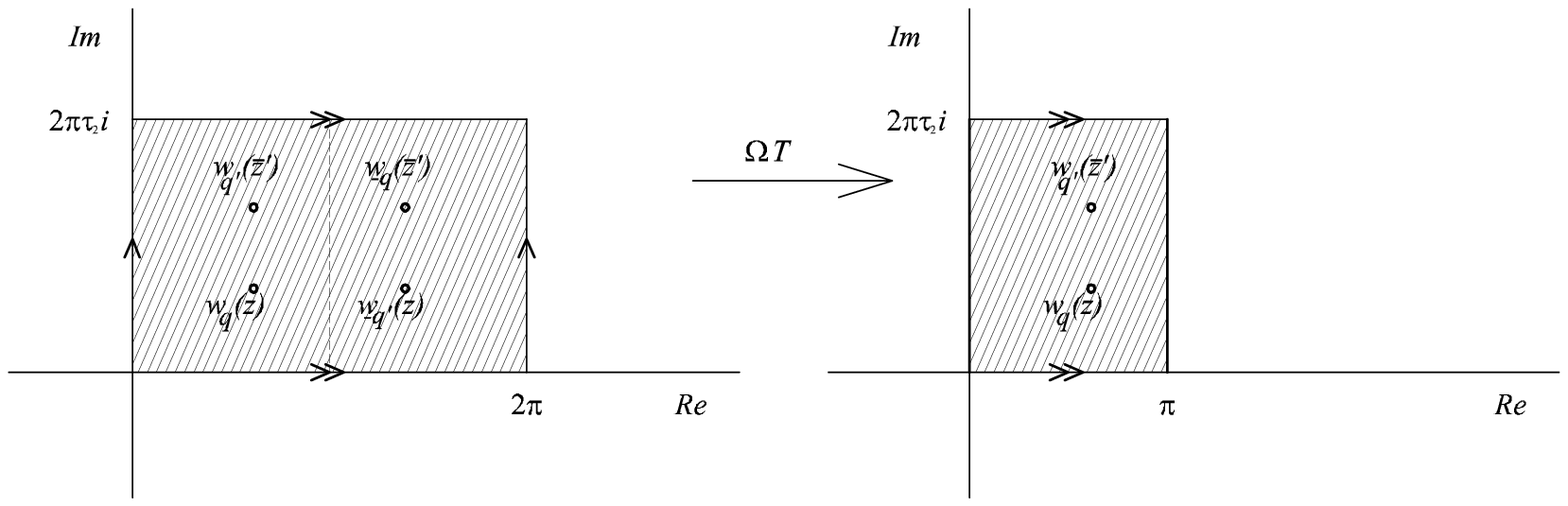}{Orbifold under $\Omega T$ in the presence of 4 Wilson
lines. The new $2d$ boundary is
$\Sigma_{o\frac{1}{2}}=S^2/\Omega=C_2$. The allowed
charges are $q=kn/4$.}{fig.t2idpt1}

This last construction corresponds to open oriented strings with {\bf
Dirichlet} boundary conditions.  The charge spectrum is $q=kn/4$,
corresponding to the winding number in string theory, and the monopole
induced processes are present allowing the \textit{gluing} of Wilson
lines with opposite charges.

\subsubsection{M\"{o}bius Strip}

Let us proceed to the parity $\tilde{\Omega}$ as given by (\r{omegat}).
The results are pictured in figure~\r{fig.t2idm2a} and are fairly
similar. Note that it corresponds to two involutions of the torus with 
$\tau=2i\tau$, one given by $\Omega$ resulting in the annulus, and
$\tilde{a}$ which maps the annulus into the M\"obius strip. Then,
for each insertion in the strip it is necessary to exist four in the torus.

\figh{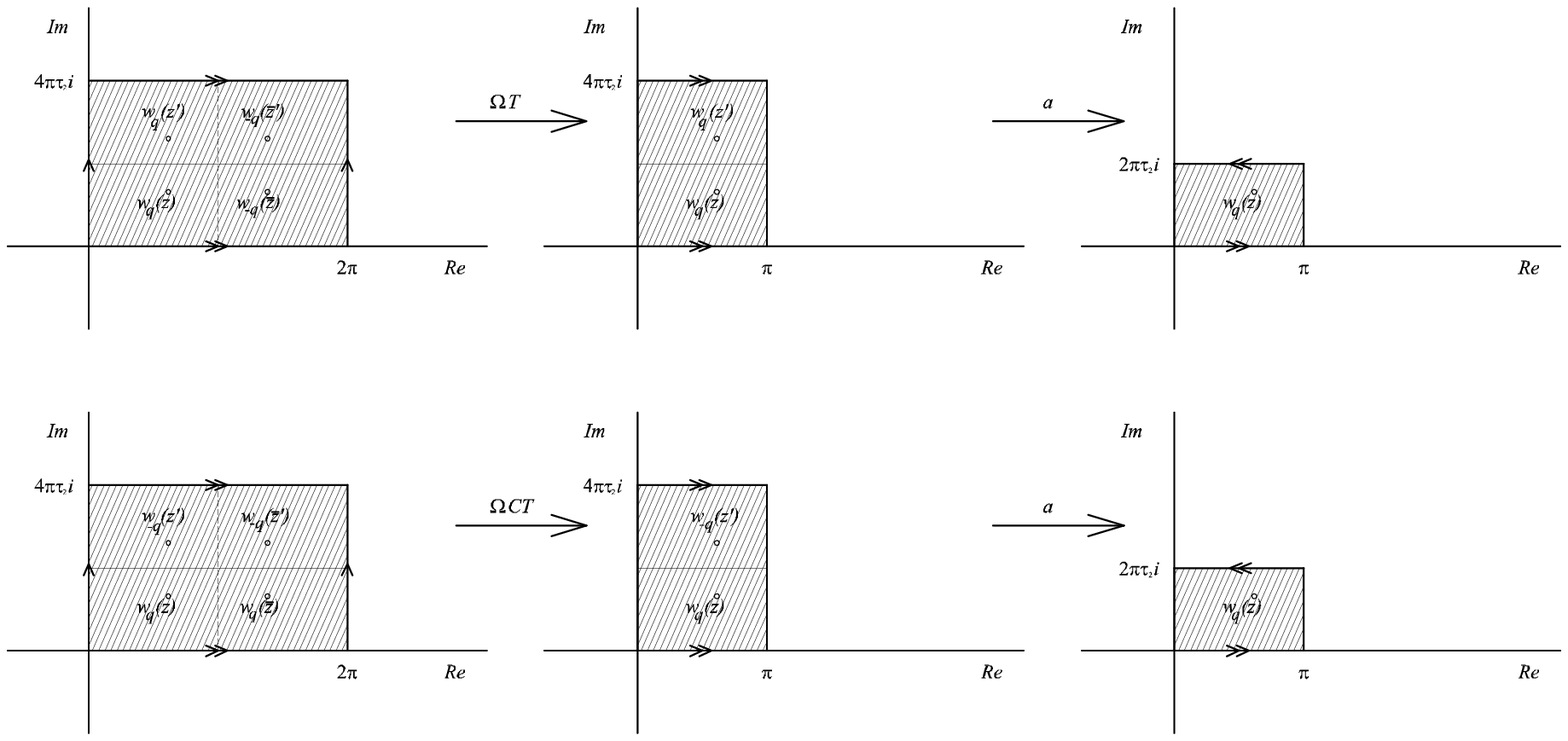}{Orbifold under $\tilde{\Omega}T$ and $\tilde{\Omega}CT$
of the torus with $\tau=2i\tau_2$ in the presence of four Wilson lines.
For $\tilde{\Omega}CT$ the Wilson lines may pierce the manifold in the
real axis and the allowed charge is $q=m$. For $\tilde{\Omega}T$ it is
not allowed the existence of boundary insertions and the admissible
charges are $q=kn/4$. The relation $z'=z+2\pi(1/2+i\tau_2)$ must
hold.}{fig.t2idm2a}

Once more we have for $\tilde{\Omega}CT$ that $B=-B=0$ in the
boundaries and $q$ is identified with $\bar{q}$ demanding the charges
to be $q=m$, which correspond to the KK momenta of string theory.
Due to this fact the monopole processes are
suppressed in the configurations allowing this kind of orbifolding.
This corresponds to {\bf Neumann} boundary conditions.

For the $\tilde{\Omega}T$ case we have the
identification of $q$ with $-\bar{q}$ demanding the charges to be $q=kn/4$,
the winding number of string theory. This time  not allowing
the monopole processes to play an important role, the charges are purelly 
\textit{magnetic}. This corresponds to {\bf Dirichlet} boundary conditions.

As discussed in subsection~\r{sec:torus_inv} we can also consider the
involution of the torus, with moduli $\tau=1/2+i\tau_2$ under $\Omega
T$ or $\Omega CT$. In this case four insertions in the torus
correspond to two insertions in the strip as presented in
figure~\r{fig.t2idm2b} for the $\Omega T$ case.
\figh{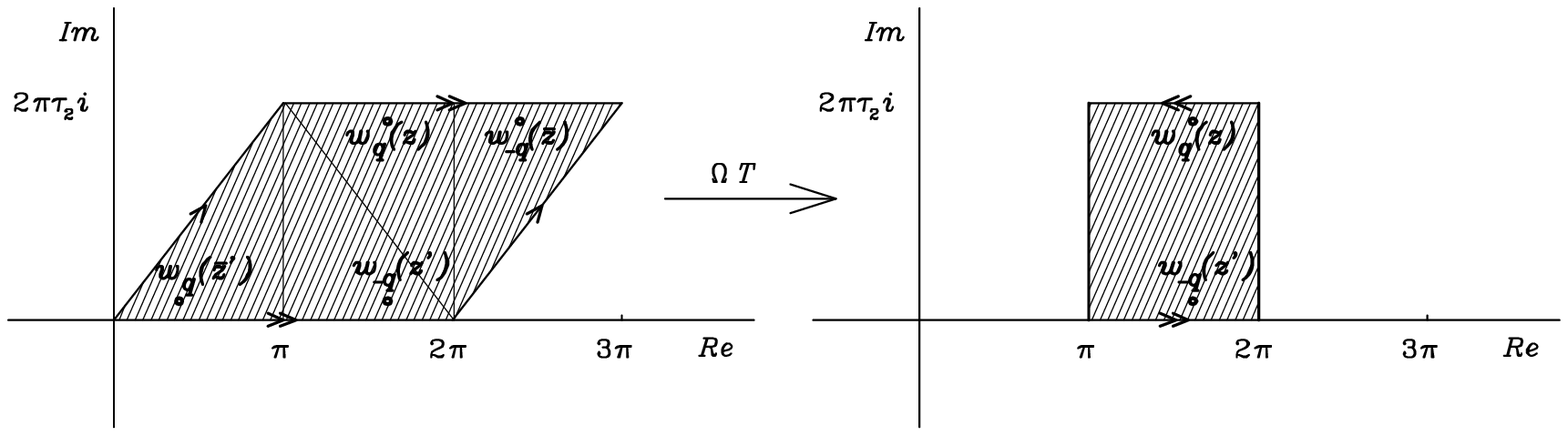}{Orbifold under $\Omega T$ of the torus with
$\tau=1/2+i\tau_2$ in the presence of four Wilson lines.}{fig.t2idm2b}

As previously explained both constructions result in the same region
of the complex plane. Note that the resulting area in both cases is
$2\pi^2\tau_2$ and that in both cases the region $[0,\pi]\times
i[0,2\pi\tau_2]$ is identified with the region $[\pi,2\pi]\times
i[0,2\pi\tau_2]$.

\subsubsection{Klein Bottle}

Finally using the parity $\Omega'$ as given by (\r{omega'}), we identify points
under $t'=1-t$, $z'=-\bz+2\pi i\tau_2$ and $\bz'=-z+2\pi i\tau_2$.
Upon orbifolding the new boundary of TM is a Klein bottle.

Again, for $\Omega'CT$, we obtain $q=m$ because $q\cong\bar{q}$. The
minimum number of insertions is two corresponding to four Wilson lines 
in the bulk. This construction corresponds to {\bf untwisted}
unoriented closed strings with only KK momenta in the spectrum.
The monopole processes are suppressed.

For $\Omega'T$ case we have $q=kn/4$ due to $q\cong-\bar{q}$.
We can have one single insertion in the bulk corresponding to two
Wilson lines or two corresponding to four Wilson lines. This
construction corresponds to {\bf twisted}
unoriented closed strings with only winding number. The monopole
processes are present and are crucial in the construction.

\figh{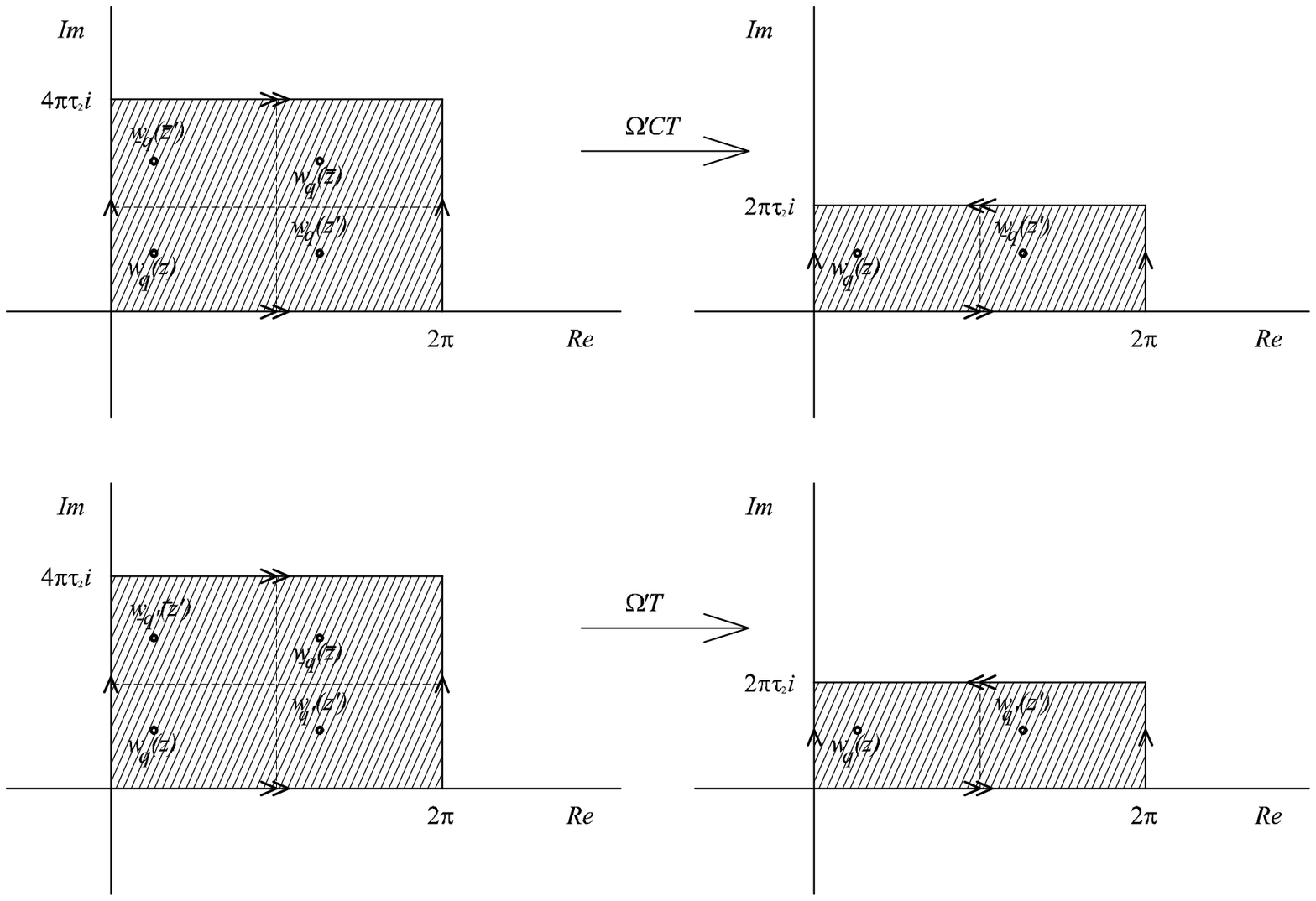}{Orbifold of $\Sigma_{\frac{1}{2}}=T^2$ under
$\Omega'CT$ and $\Omega'T$.}{fig.t2id3}

Two examples corresponding to four Wilson lines are pictured
in figure~\r{fig.t2id3}.

\subsection{Note on Modular Invariance and the Relative Modular Group\lb{sec:tmgt.mod}}

Modular invariance is a fundamental ingredient in string theory
which makes closed string theories UV finite. What about the
orbifolded theories? It is much more tricky. Each separated sector of
open and unoriented theories is clearly not invariant under a modular
transformation. The transformation $\tau\to-1/\tau$ can be interpreted
as the exchange of the holonomy cycles $\alpha$ and $\beta$ of the
torus as represented in figure~\r{fig.mod}.  Equivalently it swaps
$PT$ and $PCT$ orbifold types. But then, if the first orbifold
corresponds to the twisted sector (closed unoriented strings) or to
Dirichlet boundary conditions (open strings), the second orbifold will
correspond to the untwisted sector (closed unoriented strings) or to
Dirichlet boundary conditions (open strings). Note the sign of the
charges in figure~\r{fig.mod} representing the Klein bottle
projection.
\figh{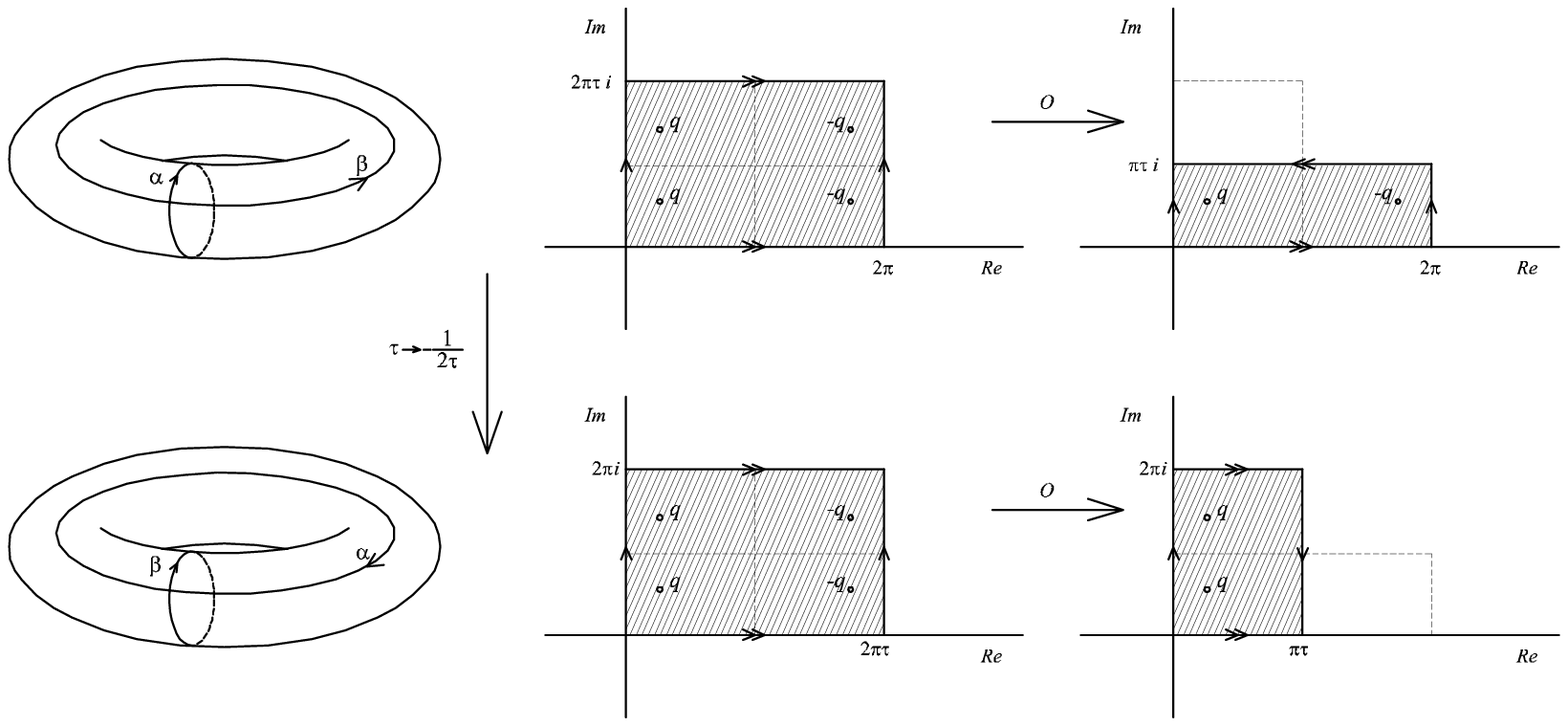}{A suitable modular transformation takes us from one
projection $PT$ (orbifold) to the other one $PCT$.}{fig.mod}

So if we actually want to ensure modular invariance we need to build a 
projection operator which ensures it. A good choice would be
\be
O=\frac{2+PT+PCT}{4}
\ee
such that the \textit{exchange} of orbifolds doesn't change
it. This fact is well known in string theory (see~\cite{P_1} for details).

In the case when we are dealing with orbifolds which result in open
surfaces the modular transformation $\tau\to-1/\tau$, according to the
previous discussion, exchanges the boundary conditions
(Neumman/Dirichlet). Note that orbifolding the target space in string
theory (or equivalently the gauge group in TMGT) is effectively
creating an orientifold plane where the boundary conditions must be
Dirichlet (as for a D-brane). This is the equivalent of
\textit{twisting} for open strings. In terms of the bulk the modular
transformation is exchanging the projections $PCT\lra PT$.

Let us put it in more exact terms. Consider some discrete group $H$ of
symmetries of the target space (or equivalently the gauge group of
TMGT). Consider now the twist by the element $h=(h_1,h_2) \in H$,
where $h_1$ twists the states in the $x_1$ direction and $h_2$ in the
$x_2$ direction.  Then the modular transformation will change the
twist as
\be
\ba{lll}
{\mathcal{T}}:&\tau\to\tau+1&(h_1,h_2)\to(h_1,h_1h_2)\vspace{.1cm}\\
{\mathcal{S}}:&\displaystyle\tau\to-\frac{1}{\tau}&(h_1,h_2)\to(h_2,h_1^{-1})
\ea
\ee

Returning to the Horava picture of describing an open string as a
\textit{thickened surface} (or double cover), in the case of the
orbifold resulting in a new open boundary the picture is similar. In
this case a modular transformation takes a open string loop, with the
ends attached to the boundaries (direct-channel picture) to a closed
string propagating from boundary to boundary (transverse-channel
picture) as pictured in figure~\r{fig.mod2} for the annulus.
\fig{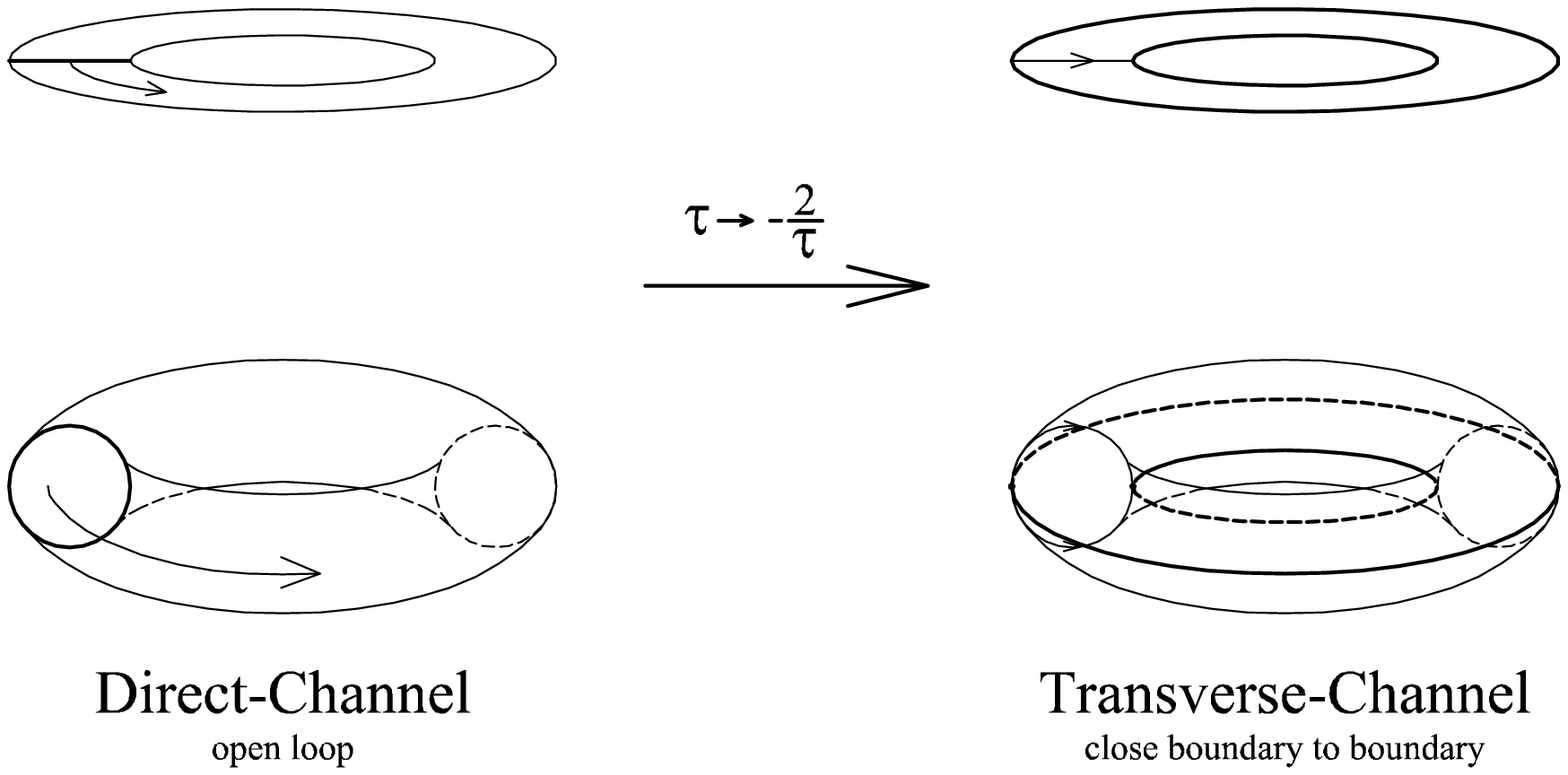}{A suitable modular transformation takes us from the
direct-channel picture to the transverse-channel picture.
Here we show this construction in terms of the thickened string}{fig.mod2}

The lower boundary of the membrane is a thickening of the string. In
the case of the open string loop we can think that the open string,
while propagating, splits into two parts.  The left modes propagate in
the top half of the torus while the right modes propagate in the
bottom half of the torus. In the case of the closed string, we again
have a splitting of the closed string exactly as before but the
propagation of the modes is transverse to the previous case as
pictured in figure~\r{fig.mod2}.

Basically this discussion explains relations~(\r{tr_BC}).
The  direct channel-picture on the disk correspond to
$\Tr_{\rm open}\left(e^{-H_o\tau}\right)$ where the trace is
considered over the possible Chan-Paton factors carried by the
open string. The  transverse-channel picture corresponds to
$\left<B\left|e^{-H_o\tilde{\tau}}\right|B\right>$
where $\left|B\right>$ stands for the states of the closed string.

So far we have concentrated on one loop amplitudes only, i.e. genus 1
world-sheet surfaces orbifolds. For the pure bosonic case this is
sufficient, but once we introduce fermions and supersymmetry new
constraints emerge at two loop amplitudes. Specifically the modular
group of closed Riemann surfaces at genus $g$ is $SL(2g,\mathbb{Z})$,
upon orbifolding there is a residual conformal group, the so called
\textbf{Relative Modular Group}~\cite{SBH_06} (see
also~\cite{SBH_06a,SBH_06b,SBH_06c}). For genus 1 this group
is trivial but for higher genus it basically mixes neighboring tori,
this means it mixes holes and crosscaps
(note that any surface of higher genus can be obtained from sewing
genus 1 surfaces). Furthermore, the string amplitudes defined on these
genus 2 open/unoriented surfaces must factorise into products of
genus 1 amplitudes. For instance a 2 torus amplitude can be thought as two
1 torus amplitudes connected trough an open string.
For a discussion of the same kind of constraints for
closed string amplitudes see~\cite{MI_01,MI_02,MI_03,MI_04,MI_05}.

The factorization and modular invariance of open/unoriented superstring
theories amplitudes will induce generalized GSO projections ensuring the
consistency of the resulting string theories.

The correct Neveu-Schwarz (NS - antiperiodic conditions, target spacetime
fermions) and Ramond (R - periodic conditions, target spacetime bosons)
sectors were built from TMGT in~\cite{TM_11}. There the minimal
model given by the coset
$M_k=SU(2)_{k+2}\times SO(2)_2/U(1)_{k+2}$ with the CS action
\be
S^{N=2}[A,B,C]=kS_CS^{SU(2)}[A]+2S_CS^{SO(2)}[B]-(k+2)S_CS^{U(1)}[C]
\ee
was considered.
It induces, on the boundary, an $N=2$
Super Conformal Field Theory (see also~\cite{TM_08} for $N=1$ SCFT).
The boundary states of the $3D$
theory corresponding to the NS and R sectors are
obtained as quantum superpositions of the 4 possible ground states
(wave functions corresponding to the first Landau level
- the ground state is degenerate) 
of the gauge field $B$, that is to say we need to choose the
correct basis of states.
The GSO projections emerge in this way as some particular
superposition of those 4 states at each boundary (for further
details see~\cite{TM_11}). It still remains to see how these
constraints emerge from genus 2 amplitudes from TM and its orbifolds.
We will discuss in detail these topics in some other occasion.

\subsection{Neumann and Dirichlet World-Sheet Boundary Conditions,\\ Monopoles Processes and Charge Conjugation}

It is clear by now that the operation of charge conjugation $C$ is selecting
important properties of the \textit{new} gauged theory. And here we
are referring to the properties of the $2D$ boundary string theory.
Gauging $PCT$ results in having an open CFT with
Neumann boundary conditions while, gauging $PT$ results in having
Dirichlet boundary conditions. So $C$ effectively selects the kind
of boundary conditions!
In the case that $PCT$ gives a closed unoriented manifold,
we obtain an untwisted theory, while $PT$ gives a twisted theory
(orientifold $X\cong-X$).
Again $C$ effectively selects the theory to be twisted or not. These
results are summarized in table~\r{tabbc}.

\bt{|cccc|cccc|}
\hline  &$P_1$  &$P_2$          &\ \ \ \ &        &$\Omega$  &$\tilde{\Omega}$&$\Omega'$\\\hline
&&&&&&&\\
$S^2\to$&$D_2$     &$RP_2$       &        &$T^2\to$&$C_2$      &$M_2$                      &$K_2$\\
&O/O     &C/U          &        &     &O/O        &O/U&C/U\\
&&&&&&&\\
$CT$&$N$&Untwisted            & &&$N$&$N$&Untwisted\\
    &q=m&q=m                  & &&$q=m$&$q=m$&q=m\\
&&&&&&&\\
$T$ &$D$&Twisted              & &&$D$&$D$&Twisted\\
    &$q=kn/4$&$q=kn/4$        & &&$q=kn/4$&$q=kn/4$&$q=kn/4$\\
&&&&&&&\\\hline
\et{\lb{tabbc}Boundary conditions and twisted sectors.}

Although these facts are closely related with strings T-duality,
the $C$~operation does not give us the dual spectrum. Upon
gauging the full theory it is only selecting the Kaluza-Klein
momenta or winding number as the spectrum of the configurations
being gauged.

From the point of view of the bulk theory the gauged configurations
corresponding to Neumann boundary conditions correspond to two Wilson
lines with one end attached to the $1D$ boundary of the new $2D$
\textit{boundary} of the membrane at $t=1/2$ and the other
end attach to the $2D$ boundary at $t=0$. For Dirichlet boundary
conditions there is one single Wilson line with both ends in the
$2D$ membrane boundary at $t=0$ and a monopole insertion in the
bulk of the $2D$ boundary at $t=1/2$. Note that the Wilson lines
do not, any longer, have a well defined direction in
time, we have gauged time inversion. These results are presented in
figure~\r{fig.monopoleDN}.
\fig{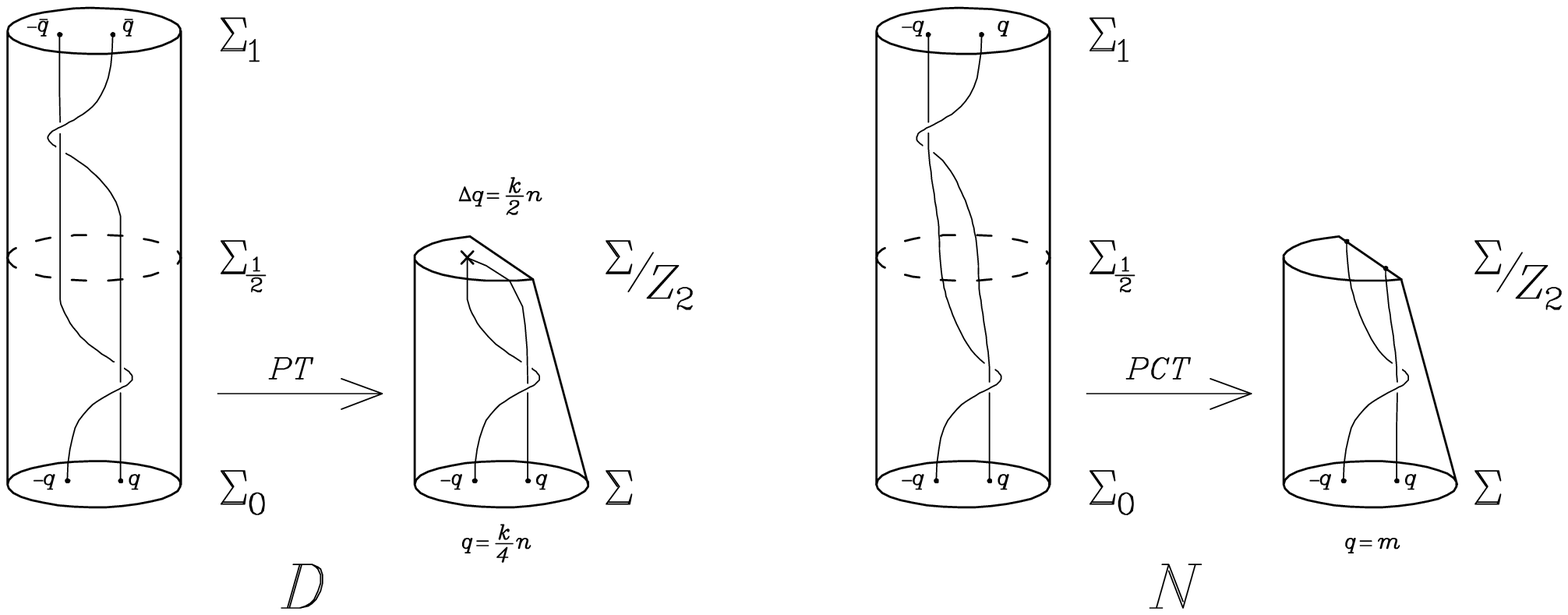}{For Dirichlet boundary conditions two Wilson lines carrying
charges $q$ and $-q$ meet in a monopole at the orbifold singular point $t=1/2$.
While for Neumann boundary conditions the two Wilson lines end in the boundary of $\Sigma/Z_2$.}{fig.monopoleDN}

For the case where we get unoriented manifolds the picture is quite
similar. There are always an even number of bulk insertions.
In the case of $PCT$ the Wilson lines which are identified
have the same charge, therefore there are no monopole processes
involved. The two Wilson lines are \textit{glued} at $t=1/2$ becoming
in the orbifolded theory one single line which has both ends attached
to $\Sigma_0$ and one point in the middle belonging to
$\Sigma_{1/2}$. In the boundary CFT we see two vertex insertions with
opposite momenta. This construction corresponds to untwisted string theories
since the target space coordinates (corresponding to the gauge
parameter $\Lambda$ in TM(GT)) are not orbifolded.

In the case of $PT$ the identification is done between charges of
opposite signs. Then two Wilson lines become one single line with its
ends attached to $\Sigma_0$, but at one end they have a $q$ charge and
in the other end they have a $-q$ charge. In $\Sigma_{1/2}$ there is
a monopole insertion which exchanges the sign of the charge. This
construction corresponds to twisted string theories since the target
space coordinates are orbifolded ($\Lambda\cong-\Lambda$).

As a final consistency check in $PCT$ the charges are always restricted
to be $q=m$ due to compatibility with the orbifold construction. By
restricting the spectrum to these form we are actually eliminating the
monopole processes for this particular configurations!

\subsection{T-Duality and Several U(1)'s}

The well know Target space or T-duality(for a review see~\cite{R_1})
of string theory is a combined symmetry
of the background and the spectrum of momenta and winding modes.
It interchanges winding modes with Kaluza-Klein modes.
From the point of view of the orbifolded TM(GT) corresponding
to open and unoriented string theories the projections $PT$
truncate the charges spectrum to $q=kn/4$ (due to demanding
$q=-\bar{q}$) which in string theory
is the winding number. The projections $PCT$ truncate the charge spectrum to
$q=m$ (due to demanding $q=\bar{q}$) which corresponds in string theory
to the KK momenta. Note that $PCT$ excludes all the monopole induced
processes while $PT$ singles out only monopole induced
processes~\cite{TM_07,TM_12,TM_14}.

T-duality is, from the point
of view of the $3D$ theory, effectively exchanging the two kinds of
projections
\be
\ba{lccc}
{\rm T-duality}:&PT&\lra&PCT\\
                &q=-\bar{q}&\lra&q=\bar{q}
\ea
\ee

This is precisely what it must do. The nature of duality in $3D$ terms
was discussed in some detail in~\cite{TM_10}. It was shown there that
it exchange topologically non trivial matter field configurations with
topologically non trivial gauge field configurations. Although charge
conjugation was not discussed there (only parity and time inversion),
this mechanism can be thought as a charge conjugation operation. Note
that $C^2=1$.

It is also rather interesting that from the point of view of the
membrane both T-duality and modular transformations are playing the
same role. In some sense both phenomena are linked by the $3D$ bulk theory.

So far we have considered only a single
compact $U(1)$ gauge group. But new phenomena emerge in the more general
case. The extra gauge sectors are necessary any how~\cite{TM_14}.

Take then the general action
with gauge group $U(1)^d\times U(1)^D$ with $d$ $U(1)$'s
noncompact and the remaining $D$'s compact.
\be
S_{d+D}=\int_Mdx^3\left[-\frac{\sqrt{-g}}{\gamma}\left(F^{M}_{\mu\nu}F_{M}^{\mu\nu} 
+F^{I}_{\mu\nu}F_{I}^{\mu\nu}\right)+\frac{\epsilon^{\mu\nu\lambda}}{8\pi}\left( 
K_{MN}A^{M}_\mu\partial_\nu A^{N}_{\lambda}+K_{IJ}A^{I}_\mu\partial_\nu A^{J}_{\lambda}\right)\right]
\label{SIJ}
\ee
where $M,N=0,\ldots,d-1$ correspond to the non compact gauge group and
$I,J=d,\ldots,d+D-1$ to the compact ones.

For a given parity $P$ we can now build an operator $O$ that acts in
every $A$ field through $PT$ and only in some of them
through $C$
\be
O=PT\left(\sum_{I'} C\delta_{I' I}+\sum_{I''}\delta_{I'' I}\right)
\ee
Due to the charges not being quantize and the non existence of
monopole-induced processes in the non compact gauge sector, the
mechanism is slightly different (see section~\r{sec:cft.bc}). But this 
operator can act as well over the noncompact sector.

For the case of open manifolds $M/PT$, $I'$ run over the indices for
which we want to impose Neumann boundary conditions
(on $\Lambda^{I'}$) and $I''$ over
the indices corresponding to Dirichlet boundary conditions.
For the case of closed manifolds $M/PT$ the picture is similar but $I'$
runs over the indices we want $\Lambda^{I'}$ to be orbifolded
(obtaining an orientifold or twisted sector).

In the case of several $U(1)$'s more general
symmetries (therefore orbifold groups)
can be considered (for instance $Z_N$).
Those symmetries are encoded in the Chern-Simons coefficient $K_{IJ}$.

\section{Conclusion and Discussion}

In this paper we have shown how one can get open and closed unoriented
string theories from the Topological Membrane. There were two major
ingredients: one is the Horava idea about orbifolding, the second is
that the orbifold symmetry was a discrete symmetry of TMGT.  The
orbifold works from the point of view of the membrane as a projection
of field configurations obeying either $PT$ or $PCT$ symmetries (the
only two kinds of discrete symmetries compatible with TMGT). For $PCT$
type projections we obtained Neumann boundary conditions for open
strings and untwisted sectors for closed unoriented strings. For $PT$
type projections we obtained Dirichlet boundary conditions for open
strings and twisted sectors for closed unoriented strings.  For $PCT$
$q=\bar{q}=m$, so only the string Kaluza Klein modes survive. In this
case the monopole induced processes are completely suppressed. For
$PT$ $q=-\bar{q}=kn/4$, so only the string winding modes survive. In
this case only monopole induced processes are present, being the
charges purelly magnetic. Charge conjugation $C$ plays an important
role in all the processes playing the role of a $Z_2$ symmetry of the
string theory target space. These results can be generalized to
symmetries of the target space encoded in the tensor $K_{IJ}$ and are
closely connected, both with modular transformations and T-duality
which exchange $PT\lra PCT$.

This work is the first part of our study of open and unoriented
string theories. In the second part~\cite{II} we shall derive the 
partition functions of the boundary CFT from the bulk
TMGT~\cite{BN_1,LR_1,O_1,AFC_1,W_2}.

Also an important issue to address in future work will be
to generalize the constructions presented here to non trivial boundary
CFT's~\cite{II}, for example WZNW models and different coset models
which can be obtained from TM with non-Abelian TMGT.

As a final remark let us note that the string photon Wilson line has
been left out. TM(GT) can take account of it as well: for any
closed $\Sigma$ there is a symmetry of the gauge group coupling tensor
$K_{IJ}\to K_{IJ}+\delta_I\chi_J-\delta_J\chi_I$ where each
$\chi_I=\chi_I[A]$ is taken to be some function of the $A^I$'s.  This
transformation affects only $B_{IJ}$ and the induced terms vanish upon
integration by parts. Once we consider the orbifold of the theory the
new orbifolded $\Sigma_o$ has a boundary and the induced terms will
not vanish any longer but induce a new action on the boundary
$\partial\Sigma_o$, they will be precisely the \textit{new} gauge photon
action of open string theories. As is well known the choice of the gauge group
of string theory, i.e. the Chan-Paton factors structure carried by
this photon Wilson line will be determined by the cancellation of
the open string theory gauge anomalies (see~\cite{P_1} and references
therein).  We postpone the proper treatment of this issue from the
point of view of TM to another occasion~\cite{II}.

\acknowledgments
The authors would like to thank
Bayram Tekin for discussions and suggestions concerning discrete
symmetries, Augusto Sagnotti for discussions and suggestions
concerning modular invariance and to Alexander Nichols and Dave
Skinner for reading the manuscript and helpful comments. PCF would
also like to thank Andr\'e Lukas and Jos\'e Nat\'ario for useful
discussions and suggestions concerning orbifold constructions and Nuno
Reis for several discussions in random topics.  The work of PCF is
supported by PRAXIS XXI/BD/11461/97 grant from FCT (Portugal).  The
work of IK is supported by PPARC Grant PPA/G/0/1998/00567 and EUROGRID 
EU HPRN-CT-1999-00161.

\end{document}